\documentclass{elsart}

\usepackage[utf8]{inputenc}
\usepackage[dvips]{graphicx}
\usepackage{amscd}
\usepackage{amsmath}
\usepackage{amssymb}
\usepackage{amstext}
\usepackage{url}
\usepackage{epsfig}
\usepackage{wasysym}
\usepackage{fancyvrb}
\usepackage{alltt}
\usepackage{hhline}
\usepackage{psfrag}
\usepackage{array}
\usepackage{setspace}
\usepackage{float}

\usepackage[sort&compress]{natbib} %sort&compress=> trier la bilblio dans un 
\def\R{\mathbb{R}} \def\T{\mathbb{T}}

\def\dg{\boldsymbol{\delta}}
\def\pg{\boldsymbol{p}}
\def\qg{\boldsymbol{q}}

\def\etz{\boldsymbol{\eta}_0}
\def\xiz{\boldsymbol{\xi}_0}

\graphicspath{{/EPSF/}{figures/}{./}}

\begin{document}

\begin{frontmatter}

  \title{Global dynamics of high area-to-mass ratios GEO space debris by means of the MEGNO indicator}
  \author{S. Valk},~\author{N. Delsate\corauthref{cor}}\ead{nicolas.delsate@fundp.ac.be}~\author{\and A. Lema\^{i}tre \and T. Carletti}
%  \author{S. Valk\corauthref{cor}}
%  \ead{stephane.valk@fundp.ac.be},~\author{N. Delsate \and A. Lema\^{i}tre \and T. Carletti}
  \address{University of Namur (FUNDP), Département de Mathématique, Unité de
    Systèmes Dynamiques, 8, Rempart de la Vierge, B-5000, Namur, Belgium}
  \corauth[cor]{Corresponding author}

  %\subtitle{Global dynamics investigations by means of the MEGNO indicator}
  %%%%%%%%%%%%%%%%%%%%%%%%%%%%%%%%%%%%%%%%%%%%%%%%%%%%%%%%%%%%%%%%%%%%%%%%
  
  \begin{abstract}
    In this paper we provide an extensive analysis of the
    global dynamics of high-area-to-mass ratios geosynchronous (GEO) space
    debris, applying a recent technique developed by~Cincotta and Sim\'{o} 
    (2000), {\it Mean Exponential Growth factor of Nearby Orbits}
    (\texttt{MEGNO}), which provides an efficient tool to investigate both
    regular and chaotic components of the phase space.
    
    We compute a stability atlas, for a large set of near-geosynchronous
    space debris, by numerically computing the \texttt{MEGNO} indicator, to
    provide an accurate understanding of the location of stable and
    unstable orbits as well as the timescale of their exponential
    divergence in case of chaotic motion. The results improve the analysis
    presented in Breiter et al. (2005) notably by considering the
    particular case of high-area-to-mass ratios space debris. The results
    indicate that chaotic orbits regions can be highly relevant, especially
    for very high area-to-mass ratios.
    
    We then provide some numerical investigations and an analytical theory
    that lead to a detailed understanding of the resonance structures
    appearing in the phase space. These analyses bring to the fore a relevant
    class of secondary resonances on both sides of the well-known pendulum-like
    pattern of geostationary objects, leading to a complex dynamics.
  \end{abstract}
  
  %%%%%%%%%%%%%%%%%%%%%%%%%%%%%%%%%%%%%%%%%%%%%%%%%%%%%%%%%%%%%%%%%%%%%%%%
  \begin{keyword}
    Solar radiation pressure -- Space debris -- \texttt{MEGNO}
    -- Detection of chaos -- Long-term evolution -- Geosynchronous orbit
    -- High area-to-mass ratios -- Secondary resonances
  \end{keyword}
\end{frontmatter}

%%%%%%%%%%%%%%%%%%%%%%%%%%%%%%%%%%%%%%%%%%%%%%%%%%%%%%%%%%%%%%%%%%%%%%%%%
%%%%%%%%%%%%%%%%%%%%%%%%%%%%%%%%%%%%%%%%%%%%%%%%%%%%%%%%%%%%%%%%%%%%%%%%%
\section{Introduction}
Recent optical surveys in high-altitude orbits, performed by the
European Space Agency 1m telescope on Tenerife (Canary islands),
have discovered a new unexpected population of 10~cm sized space
debris in near geosynchronous orbits (GEO). These objects sometimes
present highly eccentric orbits with eccentricities as high as $0.55$ 
%\citep{schildknecht04,schildknecht05}. Following the initial guess of
(Schildknecht et al., 2004 and 2005). Following the initial guess of
\citet{liouweaver04} who suggested that this new population may be 
constitued by GEO objects with high area-to-mass ratios, 
recent numerical and analytical 
investigations were performed to support this assumption
\citep{anselmo05, liouweaver05}. In addition, these authors
and others, such as \citet{chao06} and later \citet{valk07b}, presented
some detailed results concerning the short- and long-term evolution of
high area-to-mass ratios geosynchronous space debris subjected to
direct solar radiation pressure. More specifically, these latter
authors mainly focused their attention on the long-term variation of
both the eccentricity and the inclination vector. Moreover, some
studies concerning the effects of the Earth's shadowing effects on the
motion of such space debris were given in \citet{valk08}. 
However, nobody ever dealt with the question to know whether these 
orbits are really predictable or not on the time scales 
of their investigations.\\

The objective of this paper is twofold. The first goal is
the investigation of the long-term stability of high area-to-mass
ratio space debris subjected to direct solar radiation pressure, 
by means of the {\it Mean Exponential Growth factor of Nearby Orbits} 
(\texttt{MEGNO}) criterion. Second, still considering high
area-to-mass ratios, we bring to the fore a relevant 
class of additional secondary structures appearing in the phase space.\\

The paper is organized as follows. In
Section~\ref{numerical_integration}, we focus our attention to the
specification of the underlying model and we give some details about
the numerical aspects of the method. In Section~\ref{megno_section},
for the sake of completeness, we dwell upon the detailed definition of
the \texttt{MEGNO} indicator,
also providing a review of its main properties, in order to understand
the behavior of the chaos indicator.  Then in
Section~\ref{validation_section}, in the framework of the validation
of our implementation, we retrieve the results obtained by
\citet{breiter2005}. We also discuss the significance of the time of
integration, recently reported by \citet{barrio2007}. In
Section~\ref{high-area-to-mass-ratios-analysis}, we apply the
\texttt{MEGNO} technique in order to give a insightful understanding
of the stability of high area-to-mass ratio space debris. More
specifically, we show that the orbits of such peculiar space debris
are extremely sensitive to initial conditions, especially with respect
to the mean longitude and the semi-major axis. Second, we perform
extended numerical analyses, showing that the related 2-dimensional phase 
space is dominated by chaotic regions, in particular when the area-to-mass
ratio is large. In addition, we also provide some results presenting
the importance of the initial eccentricity value in the appearance of
chaotic regions. Finally, in
Section~\ref{secondary_resonances_section}, we present extensive 
numerical and analytical investigations of the additional patterns
which will be identified as secondary resonances.
%%%%%%%%%%%%%%%%%%%%%%%%%%%%%%%%%%%%%%%%%%%%%%%%%%%%%%%%%%%%%%%%%%%%%%%%%
\section{The model}\label{numerical_integration}
For the purpose of our study, we consider the modeling of a space
debris subjected to the influence of the Earth's gravity field, to
both the gravitational perturbations of the Sun and the Moon as well
as to the direct solar radiation pressure. As a consequence the
differential system of equations governing the dynamics is given by
\begin{equation*} 
  \ddot{\boldsymbol{r}} = \boldsymbol{a}_{\text{pot}} +
  \boldsymbol{a}_{\leftmoon} + \boldsymbol{a}_{\odot} +
  \boldsymbol{a}_{\text{rp}}\, ,
\end{equation*} 
where $\boldsymbol{a}_{\text{pot}}$ is the acceleration induced by the
Earth's gravity field, which can be expressed as the gradient of the 
following potential
\begin{equation}\label{app:force:earthpotential} 
U(r,\lambda,\phi) = -\frac{\mu}{r}\sum_{n=0}^{\infty}\sum_{m=0}^{n}
\left(\frac{R_e}{r}\right)^n\,\mathcal{P}_n^m(\sin\phi)(C_{nm}\cos\, m\lambda
+ S_{nm}\sin\, m\lambda)\,, 
\end{equation}
where the quantities $C_{nm}$ and $S_{nm}$ are the spherical harmonics
coefficients of the geopotential. The Earth's gravity field 
adopted is the \texttt{EGM96} model \citep{lemoine87}. In
Eq.~(\ref{app:force:earthpotential}), $\mu$ is the gravitational 
constant of the Earth, $R_e$ is the Earth's equatorial radius and the
quantities $(r,\lambda,\phi)$ are the geocentric spherical coordinates 
of the space debris. $\mathcal{P}_n^m$ are the well-known Legendre
functions. It is worth noting that the
potential of Eq.~(\ref{app:force:earthpotential}) is subsequently expressed in 
Cartesian coordinates by means of the Cunningham algorithm 
\citep{cunningham70}.\\
 
Both the accelerations $\boldsymbol{a}_{\leftmoon}$ and
$\boldsymbol{a}_{\odot}$ result from the gravity interaction with a third body 
 of mass $m_*$, where $*=\leftmoon$ and $*=\odot$, and 
can be expressed with respect to the Earth's center of mass as 
\begin{equation*}
  \boldsymbol{a}_{*} = -\mu_*
  \left(\frac{\boldsymbol{r}-\boldsymbol{r}_*}{\|\boldsymbol{r}-\boldsymbol{r}_*\|^3}  + \frac{\boldsymbol{r}_*}{\|\boldsymbol{r}_*\|^3}\right)\, ,  
\end{equation*}
where $\boldsymbol{r}$ and $\boldsymbol{r}_*$ are the geocentric
coordinates of the space debris and of the mass $m_*$, 
respectively. The quantity $\mu_*$ is the gravitational constant of
the third-body. In our implementation, we chose the high accurate
solar system ephemeris given by the Jet Propulsion Laboratory 
(\texttt{JPL}) to provide the positions of both the Sun and the Moon
\citep{standish98}.\\

Regarding direct solar radiation pressure, we assume an
hypothetically spherical space debris. The albedo of the Earth is
ignored and the Earth's shadowing effects are not taken into account
either. The acceleration induced by direct solar radiation pressure 
is given by 
\begin{equation*}\label{ch:b:rp} 
\mathbf{a_{rp}} = C_r \, P_r \left[ \frac{a_\odot}{\| \mathbf{r-r_\odot} \|}
\right]^2 \frac{A}{m} \, \frac{ \mathbf{r-r_\odot} }{\|
\mathbf{r-r_\odot} \|}\,,
\end{equation*}
where $C_r$ is the adimensional reflectivity coefficient (fixed to $1$
further on in this paper) which depends on the optical properties of
the space debris surface; $P_r=4.56 \times 10^{-6}$~N/m$^2$ is the
radiation pressure for an object located at the distance of 1~AU;
$a_{\odot}=1$~AU is a constant parameter equal to the mean distance
between the Sun and the Earth and $\boldsymbol{r}_{\odot}$ is the
geocentric position of the Sun. Finally, the coefficient $A/m$ is the
so-called area-to-mass ratio where $A$ and $m$ are the effective
cross-section and mass of the space debris, respectively.
%%%%%%%%%%%%%%%%%%%%%%%%%%%%%%%%%%%%%%%%%%%%%%%%%%%%%%%%%%%%%%%%%%%%%%%%%
\section{The Mean Exponential Growth factor of Nearby Orbits}
\label{megno_section}
For the sake of clarity we present in this section the definition and some
properties of the \texttt{MEGNO} criterion.\\

Let $\mathcal{H}(\boldsymbol{p,q})$, with $\boldsymbol{p} \in
\mathbb{R}^n$, $\boldsymbol{q} \in \mathbb{T}^n$, be a $n$-degree of
freedom Hamiltonian system and let us introduce the compact notation
$\boldsymbol{x} = (\boldsymbol{p,q}) \in \mathbb{R}^{2n}$ as well as 
$\boldsymbol{f} = (-\partial \mathcal{H}/\partial
\boldsymbol{q},\partial \mathcal{H}/\partial \boldsymbol{p}) \in
\mathbb{R}^{2n}$, then the  dynamical system is described by the
following set of ordinary differential equations
%%%%%%%%%%%%%%%%%%%%%%%%%%%%%%%%%%%%%%%%%%%%%%%%%%%%%%%%%%%%%%%%%%%%%%%%%
%%%%%%%%%%%%%%%%%%%%%%%%%%%%%%%%%%%%%%%%%%%%%%%%%%%%%%%%%%%%%%%%%%%%%%%%%
\begin{equation}
\label{equations_of_motion}
\frac{d}{dt} \boldsymbol{x}(t) =
\boldsymbol{f}(\boldsymbol{x}(t),\boldsymbol{\alpha})\,, \qquad
\boldsymbol{x} \in \mathbb{R}^{2n}\,,
\end{equation}
where $\boldsymbol{\alpha}$ is a vector of parameters entirely defined
by the model. Let $\phi(t) = \phi(t;\boldsymbol{x}_0,t_0)$ be a
solution of the flow defined in Eq.~(\ref{equations_of_motion}) with initial
conditions $(t_0,\boldsymbol{x}_0)$, then it has associated the Lyapunov
Characteristic Number (hereafter LCN), defined by~\citep{benettin80a}
% ----------------------------------------------------------------------
\begin{equation}
\label{lcn}
\lambda = \lim_{t \rightarrow \infty} \frac{1}{t}\,\ln \frac{\| \boldsymbol{\delta}_\phi(t)\|}{\|\boldsymbol{\delta}_\phi(t_0)\|},
\end{equation}
where $\boldsymbol{\delta}_\phi(t)$, the so-called {\it tangent
vector}, measures the evolution of an initial infinitesimal deviation
$\boldsymbol{\delta}_\phi(t_0) \equiv \boldsymbol{\delta}_0$ between $\phi(t)$
and a nearby orbit, and whose evolution is given by the
variational equations (terms of order
$\mathcal{O}(\boldsymbol{\delta}^2)$ are omitted)
% ----------------------------------------------------------------------
\begin{equation}
\label{variational_system}
\dot{\boldsymbol{\delta}}_{\phi} = \frac{d}{dt}\boldsymbol{\delta}_\phi(t) =
\boldsymbol{J}(\phi(t)) \, \boldsymbol{\delta}_\phi(t)\,,  
\quad  \text{with} \quad 
\boldsymbol{J}(\phi(t)) = \frac{\partial \boldsymbol{f}}{\partial
  \boldsymbol{x}}(\phi(t))\,, 
\end{equation}
where $\boldsymbol{J}(\phi(t))$ is the Jacobian matrix of
the differential system of equations, evaluated on the solution 
$\phi(t)$. Let us note that the
definition of LCN, given by Eq.~(\ref{lcn}), can also be written 
in an integral form
% ---------------------------------------------------------------------- 
\begin{equation*}
\label{lyapunov_integral} 
\lambda = \lim_{t\rightarrow\infty}\frac{1}{t} \,
\int_0^t\,\frac{\dot{\delta}_\phi(s)}{\delta_\phi(s)} \, ds\,,
\end{equation*}
where $\delta_{\phi} = \|\boldsymbol{\delta}_{\phi}\|$, $\dot{\delta}_{\phi} =
\dot{\boldsymbol{\delta}}_{\phi} \cdot \boldsymbol{\delta}_{\phi} /
\delta_{\phi}$. 
% ----------------------------------------------------------------------

The Mean Exponential Growth factor of Nearby Orbits 
$Y_\phi(t)$ is based on a modified time-weighted 
version of the integral form of LCN \citep{cincotta2000}. More
precisely
\begin{equation*}
Y_\phi(t) = \frac{2}{t} 
\, 
\int_0^t 
\, 
\frac{\dot{\delta}_\phi(s)}{\delta_\phi(s)} 
\, 
s\, ds\,,
\end{equation*}
as well as its corresponding mean value, to get rid of the
quasi-periodic oscillation possibly existing in $Y_\phi(t)$ 
% ---------------------------------------------------------------------
\begin{equation*} 
\overline{Y}_\phi(t) = \frac{1}{t} \, \int_0^t \, Y_\phi(s) \, ds\,.
\end{equation*}
% ---------------------------------------------------------------------
In the following we will omit the explicit dependence of $Y$ and
$\overline{Y}$ on the specific orbit $\phi$, when this will be 
clear from the context. 

Actually, $\overline{Y}(t)$ allows to study the dynamics for long time
scales, where generically $Y(t)$ does not 
converge, while $\lim_{t \rightarrow \infty} \overline{Y}(t)$ is well 
defined \citep{cincotta2003}. Consequently, the time evolution of
$\overline{Y}(t)$ allows to derive the possible divergence of the 
norm of the tangent vector $\boldsymbol{\delta}(t)$, giving a clear indication
of 
the character of the different orbits. Indeed, for quasi-periodic 
(regular) orbits, $Y(t)$ oscillates around the value $2$ with a linear
growth of the separation between nearby orbits. On the other hand, for chaotic
(irregular)  
motion, the norm of $\boldsymbol{\delta}$ grows exponentially with time,
and $Y(t)$ oscillates around a linear divergence 
line. \citet{cincotta2003} showed that, for the quasi-periodic orbits,
$\overline{Y}(t)$ always converges to 2, that is a fixed constant. 
Moreover, it has been shown that ordered 
motions with harmonic oscillations, i.e. orbits very close to a stable
periodic orbit, tend asymptotically to $\overline{Y}(t) = 0$.\\

These latter properties can also be used to compute efficiently a good
estimation of LCN, or similarly the Lyapunov time $T_\lambda = 
1/\lambda$, by means of a linear least square fit of
$\overline{Y}(t)$. Indeed, in the case of an irregular orbit, the time 
evolution of $\overline{Y}(t)$ may be easily written as
$$ 
\overline{Y}(t) \simeq a_\star \, t + d, \qquad t \rightarrow \infty,
$$ where $a_\star$ is simply related to LCN by the relation 
$a_\star = \lambda/2$ and $d$ is small %close to zero 
for irregular and chaotic motion. 
%But, for large $t$, $a_\star$ is much bigger than $d$ and 
Bur for regular orbits, after a transitory time, 
$d$ is not necessarily close to zero. 
Thus, the value of $d$ may be considered as the measure of how long 
the orbits sticks to a regular torus before getting chaotic 
\citep{cincotta2000}.

Regarding the numerical computation of the \texttt{MEGNO} indicator, 
we adopt the same strategy as in \citet{gozdziewski2001}. To be
specific, in addition to the numerical integrations of both the 
equations of motion and the first order variation equations, we
consider the two additional differential equations 
\begin{equation}
\label{additional_equations_megno} 
\frac{d}{dt}y = \frac{\dot{\boldsymbol{\delta}} \cdot
\boldsymbol{\delta}}{\boldsymbol{\delta} \cdot \boldsymbol{\delta}}\,,
\qquad \frac{d}{dt}w = 2\frac{y}{t}\,,
\end{equation}
which allow to derive the \texttt{MEGNO} indicators as
$$  
Y(t) = 2\,y(t)/t, \qquad \overline{Y}(t) = w(t) / t\,.
$$ The \texttt{MEGNO} criterion, unlike the common Lyapunov
variational methods, takes advantage of the whole dynamical information 
for the orbits and the evolution of its tangent vector, which results
in shorter times of integration to achieve comparable
results. Moreover, a couple of applications found in the literature
(e.g. \citealt{gozdziewski2001,gozdziewski2007};
\citealt{cincotta2000, breiter2005}) justify and confirm that 
\texttt{MEGNO} is relevant, reliable and provides an efficient way for
the investigation of the dynamics by detecting regular as well as
stochastic regimes.
%%%%%%%%%%%%%%%%%%%%%%%%%%%%%%%%%%%%%%%%%%%%%%%%%%%%%%%%%%%%%%%%%%%%%%%%%
%%%%%%%%%%%%%%%%%%%%%%%%%%%%%%%%%%%%%%%%%%%%%%%%%%%%%%%%%%%%%%%%%%%%%%%%% 
\subsection{\texttt{MEGNO} and numerical integrations}
As previously mentioned, in order to evaluate the \texttt{MEGNO}
indicator, we have to integrate the differential system of 
Eq.~(\ref{equations_of_motion}), the linear first order 
variational system of Eq.~(\ref{variational_system}), as well as 
the two additional differential Eq.~(\ref{additional_equations_megno}). 
We choose to write both
the expressions of the perturbing forces and the variational system, i.e. the 
Jacobian matrix, in rectangular coordinates {\it positions-velocities}.
%This approach has
%two advantages. First, the explicit computation of the right-hand
%sides of the variational system of equations allows to apply the
%method described in Section~\ref{megno_section}, avoiding the
%practical difficulties present in the classical method of neighboring
%trajectories (two particles method). Second, 
In such a way we can overcome both the null eccentricity and the null
inclination singularity present in the dynamics of space debris
\citep{valk07a}. Moreover, the explicit analytical expressions of the
vector fields allow us to avoid the difficulties inherent in the classical
method of neighboring trajectories (two particles method).

In order to integrate numerically the two differential systems of 
equations, we adopted the variable step size Bulirsh-Stoer algorithm
(see e.g. Bulirsh and Stoer, 1966, and Stoer and Bulirsch, 1980). Let 
%(see e.g. \citealt{bulirsh-stoera}, and \citealt{bulirsh-stoerb}). Let 
us note that, for the purpose of validation, the numerical
integrations were also made with a couple of other numerical
integrators. However, the Bulirsh-Stoer algorithm seems to be the best
compromise between accuracy and efficiency. Moreover, as quoted by
\citet{wisdom83}: {\it What is more important for this study,
\citet{benettin80a} found that the maximum LCE\footnote{Lyapunov
Characteristic Exponent.} did not depend on the
precision of their calculation. It appears likely that as long as a
certain minimum precision is kept, maximum LCE's may be accurately
computed, even though it is not possible to precisely follow a
specified trajectory for the required length of time}.

Although this latter observation was formulated in the framework of
both Lyapunov variational method and Hamiltonian systems, it seems
that it remains relevant in the computation of the \texttt{MEGNO}
criterion, at least in the particular case of our analysis.

\subsection{Influence of the initial tangent vector $\boldsymbol{\delta}_0$}
By construction \texttt{MEGNO} depends on the initial value of the tangent
vector $\boldsymbol{\delta}_0$ %. 
%However, on very large $t$ (formally for 
%$t\longrightarrow \infty$) the initial value of the tangent vector 
%not influence significantly the detection of 
%chaotic regions. Nevertheless 
as the LCE \citep{benettin80a}. That's way 
we preferred to adopt the strategy of initialize
randomly the initial tangent vectors in order to avoid some parts of the 
{\it artificially created zones of low \texttt{MEGNO} due to the proximity of
$\boldsymbol{\delta}_0$ to the minimum Lyapunov exponent direction}
\citep{breiter2005}. Moreover, as pointed out by
\citet{gozdziewski2001}, the random sampling of
$\boldsymbol{\delta_0}$ is relevant in the sense that different
initial tangent vectors can lead to different behaviors of the
\texttt{MEGNO} time evolution while considering the same orbit. This
observation has been reported in the framework of extra-solar planetary
systems and seems to remain similar in the case of Earth orbiting
objects and more generally for high-dimensional dynamical systems
(having more than 3 degrees of freedom).

Regarding the impact of the choice of the initial tangent vector
$\boldsymbol{\delta}_0$, we performed a set of exhaustive numerical
investigations of regular orbits. More specifically, we compared the
time-evolution of \texttt{MEGNO} using different initial tangent
vectors and identical generic initial conditions. The results confirm
that the random choice of the initial tangent vector induces a
significant random behavior in the way \texttt{MEGNO} approaches the
limit value $2$, hence preventing this information from being useful
to check the stability/instability character of regular
orbits. Actually, when considering a slightly perturbed two-body
problem (such as the central attraction disturbed by the oblateness of
the Earth), the way \texttt{MEGNO} converges to $2$ is completely
unpredictable  , leading to more or less 50\% of convergence of
$\overline{Y}(t)$ to 2 from above and the other remaining 50\% from
below. This result is formally discussed in the following
subsections. However, when the order of magnitude of the perturbation
is larger, the result does not completely hold anymore. In particular,
when considering the perturbing effects induced by the 1:1 resonance,
the \texttt{MEGNO} evolution no longer depends on the random choice of
the initial tangent vector. In this latter case, the intrinsic
stability of the chosen orbits seems also to dictate the evolution of 
\texttt{MEGNO} as reported in \citet{cincotta2003}. More
specifically, the stability of the orbit seems to influence the
time evolution of \texttt{MEGNO} the more the orbit is closer
to a stable or unstable equilibrium point. For instance, regarding the
orbits extremely close to a stable equilibrium point, 
\texttt{MEGNO} generally approaches slowly the limit value $2$ from
below, even though some infrequent orbits present a \texttt{MEGNO}
convergence from above. Conversely, the orbits initially close to the
separatrices generally present a \texttt{MEGNO} approaching the value
2 from above.
%%%%%%%%%%%%%%%%%%%%%%%%%%%%%%%%%%%%%%%%%%%%%%%%%%%%%%%%%%%%%%%%%%%%%%%%%
%%%%%%%%%%%%%%%%%%%%%%%%%%%%%%%%%%%%%%%%%%%%%%%%%%%%%%%%%%%%%%%%%%%%%%%%%
%% \begin{figure}[!h]
%%     \begin{tabular}{ll}
%%       \hspace{-2cm}
%%       \includegraphics[width=9cm]{convHB.eps}&
%%       \includegraphics[width=9cm]{convLenteVers2.eps}
%%     \end{tabular}
%% \caption{\label{convergence_megno} Effect of the random choice of
%% initial tangent vector $\boldsymbol{\delta}_0$ and significance of the
%% time of integration. [Left] Illustratation of different time
%% evolutions of $\overline{Y}(t)$ only when considering different choice
%% of the initial tangent vector $\boldsymbol{\delta}_0$. The propagated
%% orbits are the same in each case. [Right] Illustration of very slow
%% convergence to the limit $\overline{Y}(t)=2$ for various regular
%% orbits.}
%% \end{figure}
%------------------------------------------------------------------------
\subsection{\texttt{MEGNO} for integrable systems}
In this section we will study the \texttt{MEGNO} indicator for
integrable Hamiltonian systems and we will show that generically (if the
system is not isochronous) it always converges to $2$, moreover the
way $Y(t)$ reaches this limit value, say from higher or lower
values, depends only on the choice of the initial tangent vector and
not on the orbit itself.

So let us consider an integrable Hamiltonian system and suppose to
write it in action-angle variables, $\mathcal{H}=\mathcal{H}(\pg$), 
where $\pg \in B \subset {\R}^n$ denotes the action variables and $\qg \in
\T^n$ denotes the angle variables. Then the Hamiltonian equations are
\begin{equation*}
%  \left\{
  \begin{array}{lcl}
    \displaystyle \dot{\pg} & = & 0\,, \\
    \displaystyle \dot{\qg} &
    = & \displaystyle \frac{\partial \mathcal{H}}{\partial \pg} =
    \boldsymbol{\omega}(\pg)\,.
  \end{array}
%  \right.
\end{equation*}
The tangent space (to a given orbit) can be split into the action and angle 
direction, namely $\dg=(\dg_p,\dg_q)$, thus the variational system can
be written as 
\begin{equation*}
  \setstretch{1.4}
%  \left\{
  \begin{array}{lcl}
    \displaystyle \dot{\dg_p} & = & 0\,, \\
    \displaystyle \dot{\dg_q} & = & \displaystyle \frac{\partial^2
      \mathcal{H}}{\partial \pg^2} \dg_p = 
    M(\pg) \, \dg_p \, .
%                & = & \displaystyle \frac{\partial
%                \boldsymbol{\omega}}{\partial  \boldsymbol{p}}\,. 
  \end{array}
%  \right.
\end{equation*}

\begin{figure}[!t]
  \begin{center}
    \begin{tabular}{cc}
      \includegraphics[width=.46\textwidth]{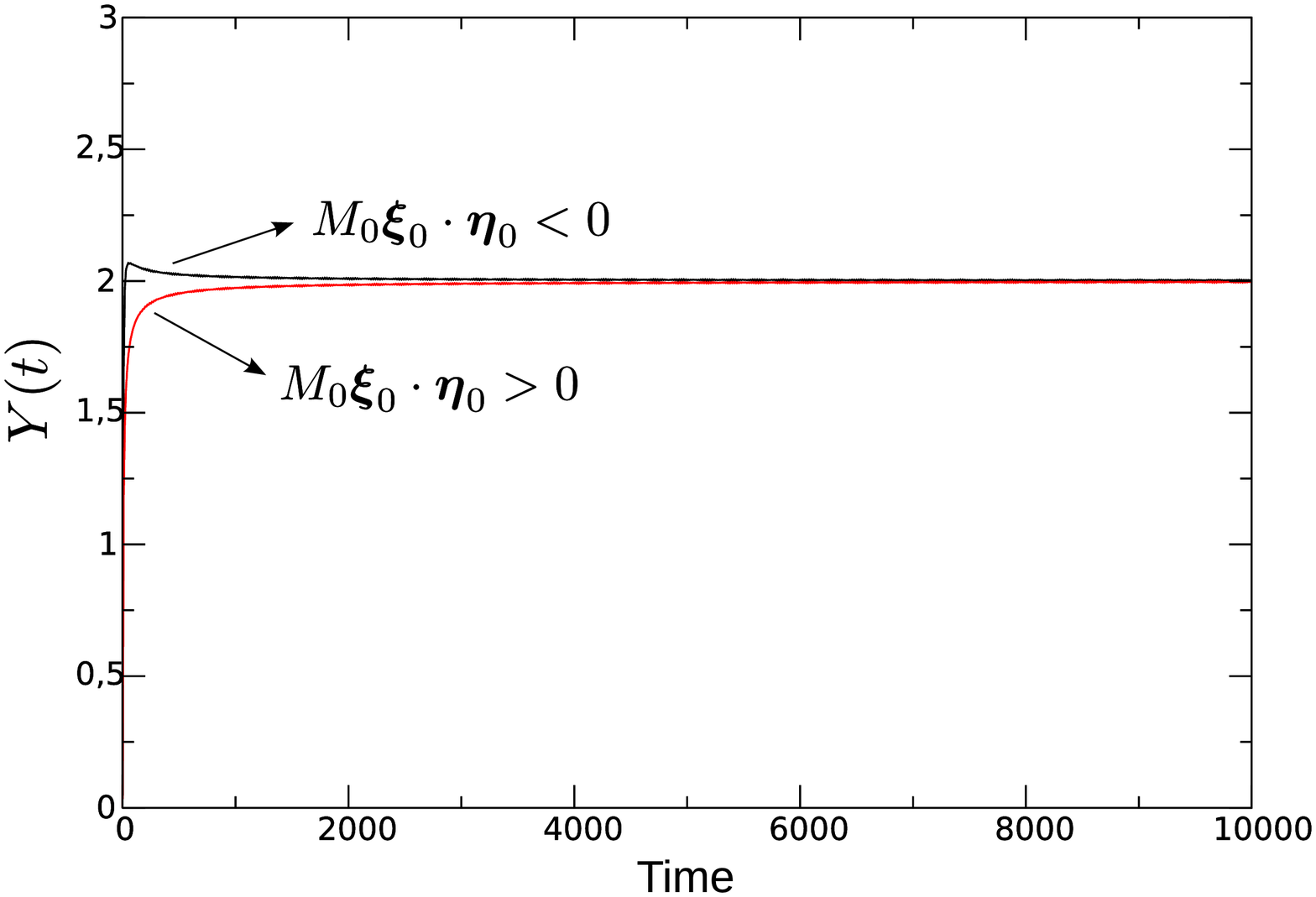}
      &
      \includegraphics[width=.46\textwidth]{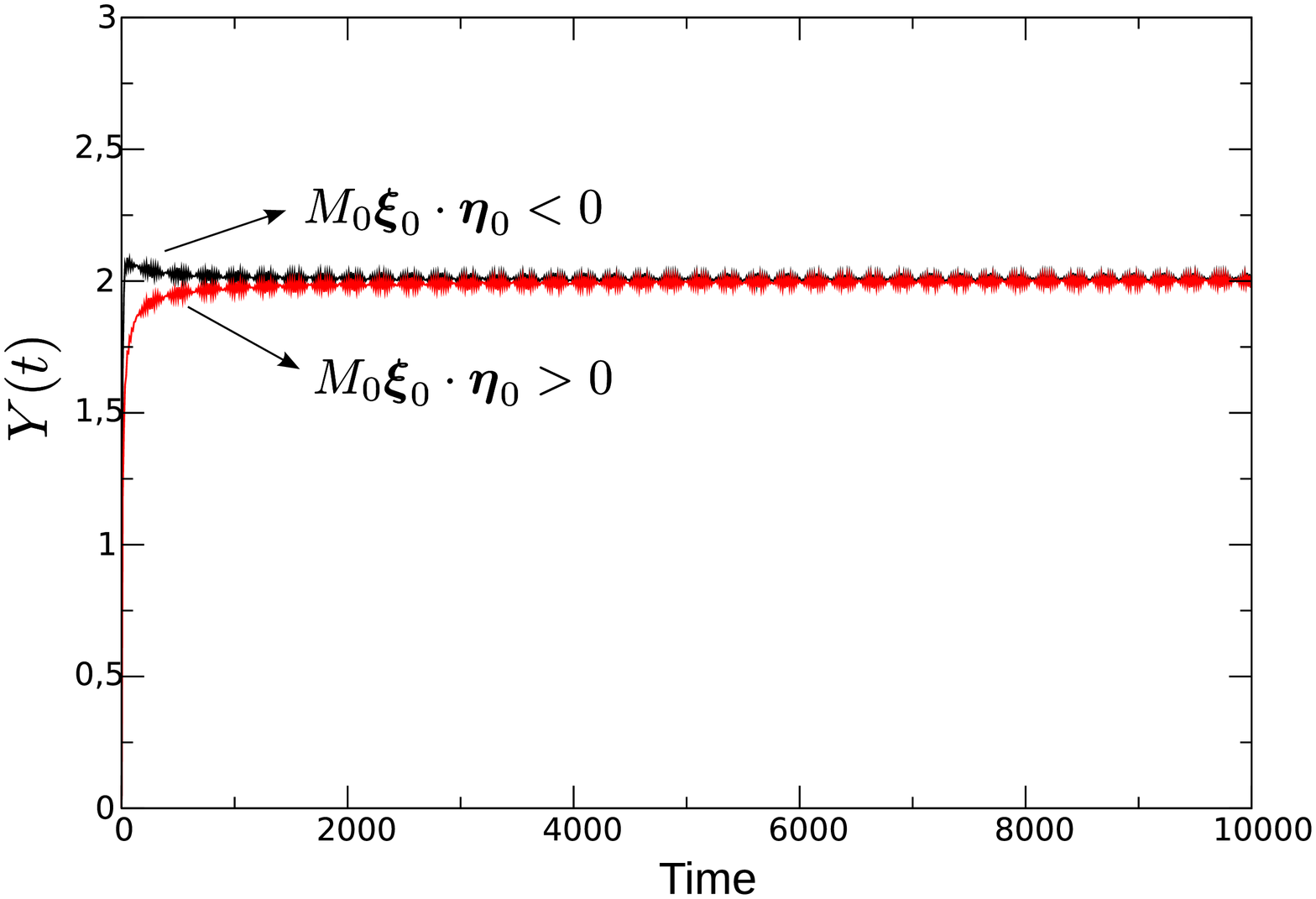} 
    \end{tabular}
  \end{center}
  \caption[]{\label{evolMegnoQuasiIntgSyst}\texttt{MEGNO} for quasi--integrable 
    adimensional Hamiltonian system. We 
    consider the evolution of $Y_{\phi_{\epsilon}}$ for the system
    $H=p_1^2/2+p_2+\epsilon \cos q_1+\epsilon \cos (q_1-q_2)$. On the left
    panel $\epsilon = 10^{-4}$, while on the right panel
    $\epsilon=10^{-3}$. In both cases $\epsilon$ is small enough to
    confirm the theoretical predictions; let observe that in this case the
    matrix $M$ is given by 
    $
    \left(
    \begin{smallmatrix} 
      1 & 0 \\ 0 & 0
    \end{smallmatrix}
    \right)
    $  and thus the sign condition reads
    $M\boldsymbol{\delta}_{p,0}\cdot
    \boldsymbol{\delta}_{q,0}=\delta_{p,0}^1\delta_{q,0}^1$. 
    %The time unit has no physical meaning.}
    The unit of time corresponding to $1/10$ of period of the orbit.}
\end{figure}

%\subsection{Isochronous systems and non-isochronous systems}
If the system is isochronous then $M\equiv 0$, thus $\dg_p$ and
$\dg_q$ are constant and $Y(t)=0$ for all $t$.  On the other hand, if the
system is 
non-isochronous we get $\dg_p(t) = \dg_p(0)$ and $\dg_q(t)=\dg_q(0) +
M(\pg(0))\,\dg_p(0)t$. To simplify the notations, 
let us introduce
$$M\left(\pg(0)\right) = M_0, \quad \dg_p(0)=\xiz \text{  and  }
\dg_q(0)=\etz\,.$$ 
Using the definition of \texttt{MEGNO}, we get
\begin{equation*}
  Y(t) = \frac{1}{t} \int_0^t \frac{(M_0\xiz)^2 s + M_0\xiz \cdot
  \etz}{(\xiz)^2+(\etz)^2+2M_0\xiz \cdot \etz s + (M_0\xiz)^2 s^2} s\; ds\,,
\end{equation*}
and this integral can be explicitly computed, obtaining 
\begin{equation}\label{megnoNonIsoSyst}
\setstretch{2}
\begin{array}{lcl}
\displaystyle   Y(t) & = & \displaystyle 2-\frac{M_0\xiz\cdot \etz}{t(M_0\xiz)^2}\log[1+2M_0\xiz \cdot
  \etz t+(M_0\xiz)^2t^2] + \\
  & - & \displaystyle \frac{2}{t} \frac{\sqrt{(M_0\xiz)^2-(M_0\xiz \cdot
  \etz)^2}}{(M_0\xiz)^2} \left[\arctan\frac{M_0\xiz \cdot \etz+(M_0\xiz)^2t^2}{\sqrt{(M_0\xiz)^2-(M_0\xiz \cdot \etz)^2}}\right.\\
  & - & \displaystyle \left.\arctan\frac{M_0\xiz \cdot \etz}{\sqrt{(M_0\xiz)^2-(M_0\xiz \cdot \etz)^2}}\right]\,.
\end{array}
\end{equation}
One can check that the square root is well defined, i.e. positive, and
thus one can cast Eq.~(\ref{megnoNonIsoSyst}) into
\begin{equation*}
  Y(t) = 2-\frac{M_0\xiz \cdot \etz}{t}F_1(t)-\frac{1}{t}F_2(t)\,,
\end{equation*}
where $F_1$ and $F_2$ are positive functions and $F_2$ is bounded. 
We can then conclude that (see Fig. \ref{evolMegnoQuasiIntgSyst})
\begin{enumerate}
  \item if $M_0\xiz \cdot \etz > 0$ then $Y(t)$ approaches $2$ from below;
  \item if $M_0\xiz \cdot \etz < 0$ then $Y(t)$ approaches $2$ from above, in
  fact for large $t$ the first contribution dominates the bounded term $F_2$. 
\end{enumerate}
%%%%%%%%%%%%%%%%%%%%%%%%%%%%%%%%%%%%%%%%%%%%%%%%%%%%%%%%%%%%%%%%%%%%%%%%%%%%%%
%\subsection{Quasi-integrable non-isochronous systems}\label{sec:qintegsys}

In this last part we will consider if and under which
assumptions the previous results concerning the
convergence $Y\rightarrow 2$ are still valid, for a quasi--integrable
Hamiltonian system of the form
$H(\boldsymbol{p},\boldsymbol{q},\epsilon)=H_0(\boldsymbol{p})+\epsilon
V(\boldsymbol{p},\boldsymbol{q})$.  The main idea is the following: fix
$\epsilon>0$, but small, and consider a \lq\lq non--chaotic\rq\rq orbit
$\phi_{\epsilon}$, namely an orbit without a positive Lyapunov
exponent (or ``if you prefer'' with a bounded \texttt{MEGNO}), then if 
$\epsilon$ is sufficiently small this orbit is a perturbation of an orbit
existing also for $\epsilon = 0$, $\phi_0$, and we can check that
$Y_{\phi_{\epsilon}}=Y_{\phi_{0}}+\mathcal{O}(\epsilon)$, hence the
smallness of such $\epsilon$--correction cannot change \lq\lq the way
$Y$ goes to~$2$\rq\rq.  More precisely, the Hamilton equations are now
\begin{equation*}
  \setstretch{1.6}
  \begin{array}{lcl}
    \displaystyle \dot{\boldsymbol{p}} 
    &=& 
    \displaystyle -\frac{\partial \mathcal{H}}{\partial
      \boldsymbol{q}}=-\epsilon \frac{\partial V}{\partial \boldsymbol{q}}\\  
    \displaystyle \dot{\boldsymbol{q}} 
    &=& \displaystyle \frac{\partial \mathcal{H}}{\partial
      \boldsymbol{p}}=\boldsymbol{\omega}(\boldsymbol{p})+\epsilon
    \frac{\partial V}{\partial \boldsymbol{p}}  
    \, ,
  \end{array}
\end{equation*}
and a similar decomposition can be done for the variational system
\begin{equation*}
  \setstretch{1.6}
  \begin{array}{lcl}
    \displaystyle \dot{\boldsymbol{\delta}}_p &=&
    \displaystyle -\epsilon \frac{\partial^2
      V}{\partial\boldsymbol{p}\partial \boldsymbol{q}}\boldsymbol{\delta}_p-\epsilon
    \frac{\partial^2 V}{\partial \boldsymbol{q}^2}\boldsymbol{\delta}_q  \\  
    \displaystyle \dot{\boldsymbol{\delta}}_q &=&
\displaystyle     \left(\frac{\partial^2 \mathcal{H}}{\partial
       \boldsymbol{p}^2}+\epsilon \frac{\partial^2
      V}{\partial\boldsymbol{p}^2}\right)\boldsymbol{\delta}_p +\epsilon \frac{\partial^2
      V}{\partial\boldsymbol{p}\partial \boldsymbol{q}} \boldsymbol{\delta}_q 
     \, .
  \end{array}
\end{equation*}
Looking for $\boldsymbol{\delta}_p$ and $\boldsymbol{\delta}_q$ as
$\epsilon$--power series,
i.e. $\boldsymbol{\delta}_p=\boldsymbol{\delta}_{p,0}+\epsilon
\boldsymbol{\delta}_{p,1}+\dots$ and
$\boldsymbol{\delta}_q=\boldsymbol{\delta}_{q,0}+\epsilon \boldsymbol{\delta}_{q,1}+\dots$,
and collecting together, in the definition of \texttt{MEGNO}, terms
contributing 
to the same power of $\epsilon$, we can thus get
\begin{equation}\label{eq:megnoqinteg}
  \setstretch{1.6}
\begin{array}{lcl}
  \displaystyle Y_{\phi_{\epsilon}}(t)&=& \displaystyle \frac{1}{t}\int_0^t \frac{(M_0\boldsymbol{\delta}_{p,0})^2s+M_0\boldsymbol{\delta}_{p,0}\cdot \boldsymbol{\delta}_{q,0}}
  {(\boldsymbol{\delta}_{p,0})^2+(\boldsymbol{\delta}_{p,0})^2+2M_0\boldsymbol{\delta}_{p,0}\cdot
  \boldsymbol{\delta}_{q,0} s+(M_0\boldsymbol{\delta}_{q,0})^2s^2}s\,
  ds+\mathcal{O}(\epsilon)\notag \\
  &=& \displaystyle Y_{\phi_{0}}(t)+\mathcal{O}(\epsilon)\, .
\end{array}
\end{equation}
%%%%%%%%%%%%%%%%%%%

\section{Validation of the method}
\label{validation_section}
To validate our method we first apply the technique on a simplified
model, containing only the Earth's gravity field expanded up to the
second degree and order harmonics, namely, $J_2 = -C_{20}, C_{22}$ and
$S_{22}$. For the purpose of the analysis, we followed a set of
12\,600 orbits, propagated over a 30~years time span, that is the
order of $10^4$ fundamental periods (1~day) empirically required by
the method \citep{gozdziewski2001}. As reported in
\citet{breiter2005}, a 30~years time span seems to be relatively
small for long-term investigations of geosynchronous space
debris. However, the numerical integration of 
variational equations in addition to the extrapolation of the orbit is
quite time consuming. Indeed, the simulation with an entry level step
size of 400~seconds takes approximately 20~seconds per orbit when
including only the Earth's gravity field, whereas it takes 42~seconds
with a complete model. Thus, the examination of large sets of initial 
conditions can take a lot of time (typically $5$ days for $10^4$~orbits). 
On the other hand, the analysis of the following section will bring to the
fore some indications about the Lyapunov times %and it will result
(smaller than 30~years). As a consequence, the integration time can be 
considered as sufficiently large in the particular case of our study.\\

For the purpose of this validation study, we consider a set of initial
conditions defined by a mean longitude $\lambda$ grid of $1^{\circ}$, spanning
$90^{\circ}$ on both sides of the first stable equilibrium point and a
semi-major axis $a$ grid of 1~km, spanning the $42164 \pm 35$~km
range. The other fixed initial conditions are $e_0 = 0.002$ for the
eccentricity, $i_0 = 0.004$~rad for the inclination, $\Omega_0 = \omega_0
=0$~rad for the longitude of the ascending node and the argument of
perigee, respectively. These values have been fixed to compare our
results for the nearly-geosynchronous orbits with the ones
of~\citet{breiter2005}. As pointed out by \citet{breiter2005}, due
to the 1:1 resonance, good variables to present our results will be
$(a_0,\sigma_0)$, where $a_0$ is the osculating initial semi-major axis and
$\sigma$ is the so-called resonant angle, i.e. $\sigma = \lambda -
\theta$, where $\theta$ is the sidereal time.\\

\begin{figure}[!t]
  \begin{center} 
    \begin{tabular}{cc}
      \includegraphics[width=.46\textwidth]{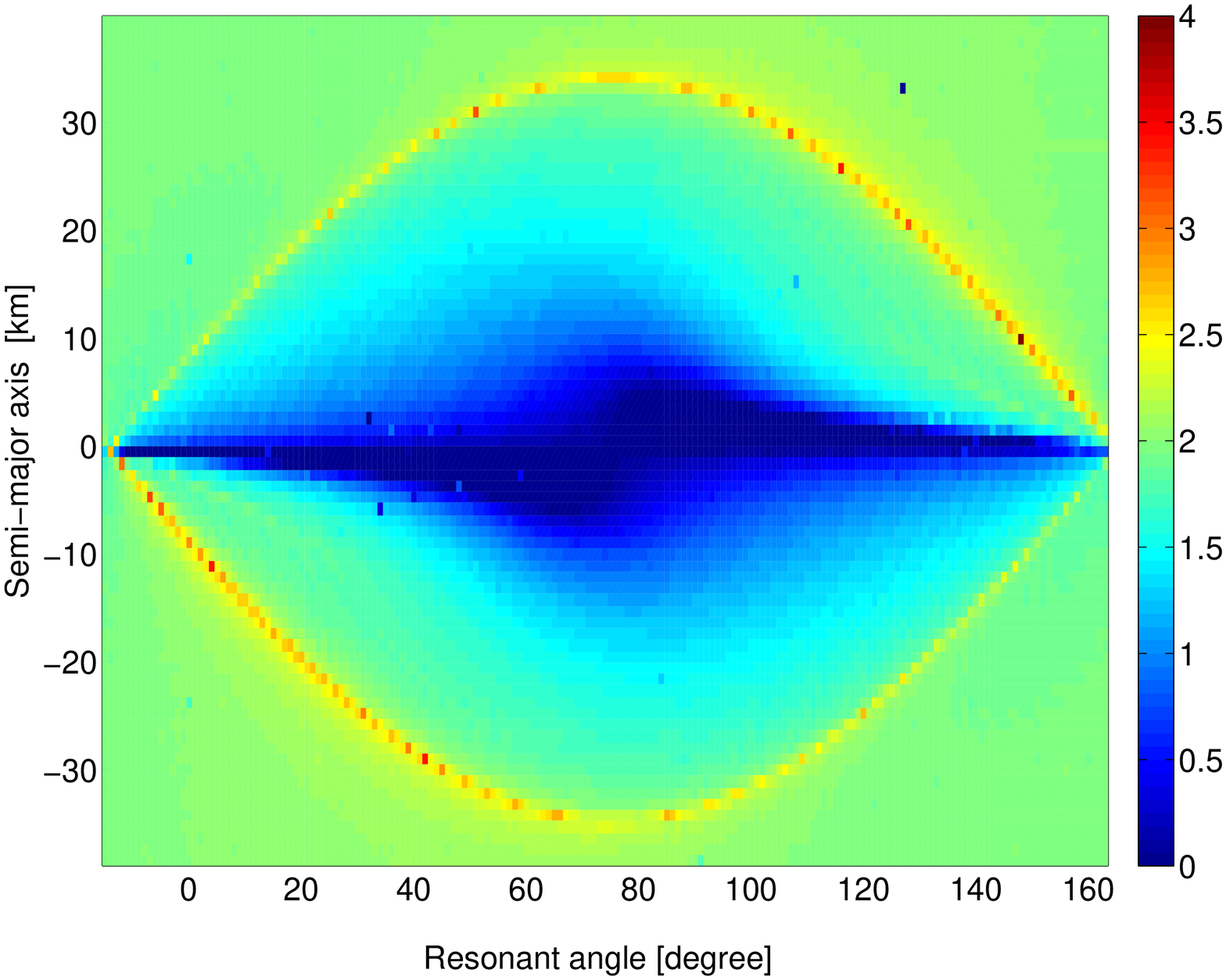} 
      &
      \includegraphics[width=.46\textwidth]{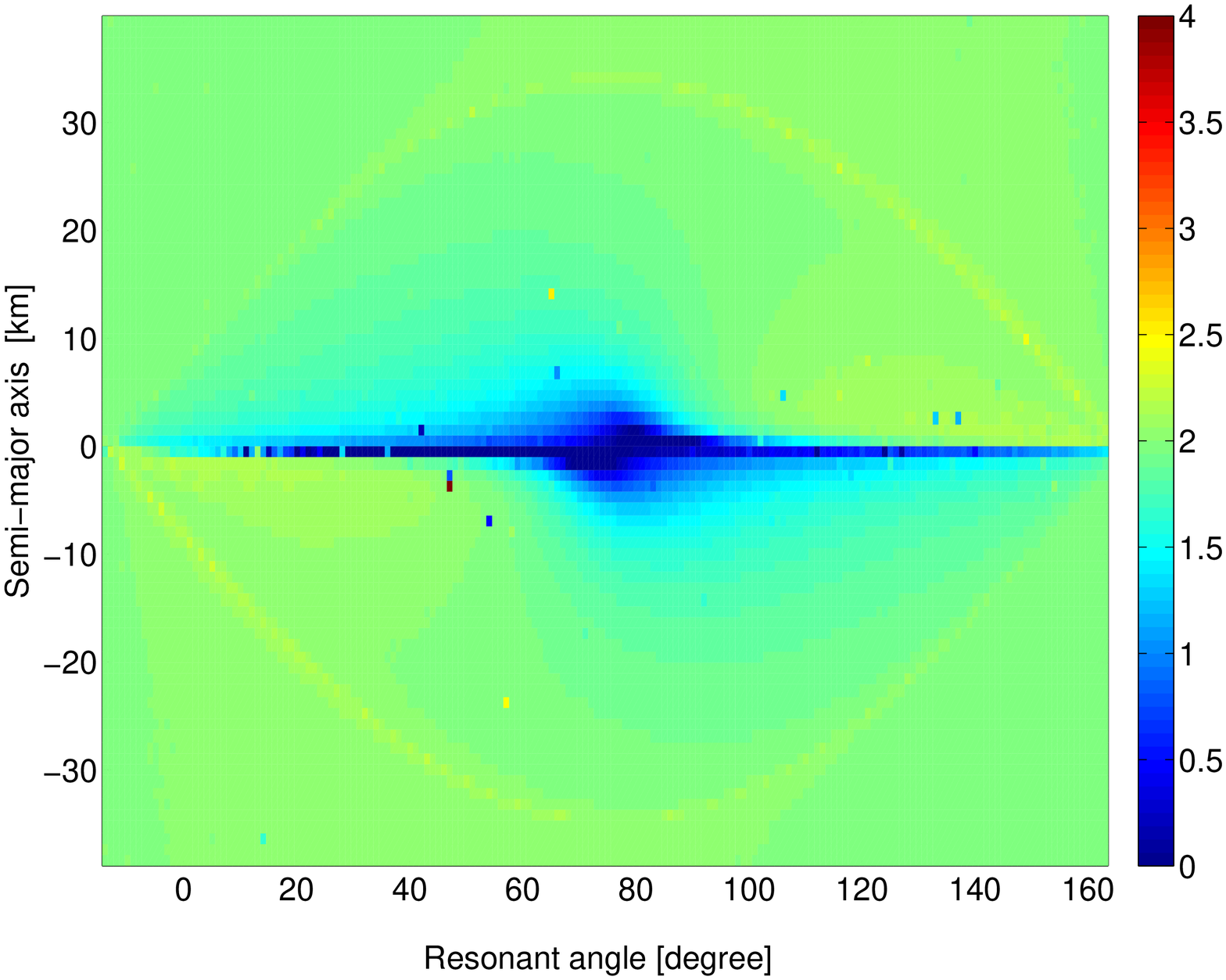} 
    \end{tabular}
    \caption{\label{samplecs22}The \texttt{MEGNO} computed as a 
    function of initial mean longitude $\lambda_0$ and osculating semi-major
    axis $a_0$. The equations of motion include the central body
    attraction as well as the second degree and order harmonics
    $J_2,C_{22}$ and $S_{22}$. The mean longitude grid is $1^{\circ}$
    and the semi-major axis grid is $1$~km, spanning the $42164 \pm
    35$~km range. The initial conditions are $e_0 = 0.002$, $i_0 =
    0.004$~rad, $\Omega_0 = \omega_0 = 0$~rad. Time at epoch is
    25~January~1991. The patterns have been obtained using two
    different times of integration, $t_f = 30$~years [left] and $t_f =
    300$~years [right].}
  \end{center}
\end{figure} 

Figure~\ref{samplecs22}~(left panel) shows the \texttt{MEGNO} values computed
using 30~years of integration time. We
identify clearly a blow-up of the typical double pendulum-like pattern
related to the 1:1 resonance. Here, we plot only over a horizontal 
range of $180^{\circ}$, i.e. only one eye. 
The existence of both the stable and the two unstable equilibrium 
points can be easily inferred. We observe that the phase 
space seems to be essentially filled in with \texttt{MEGNO} values
$\overline{Y}(t) \simeq 2$, that is plenty of regular orbits. Moreover, 
the two separatrices are also identifiable and are associated with
neighboring \texttt{MEGNO} values $2 < \overline{Y}(t) \leq 
4$. Therefore, following the properties defined in
Section~\ref{megno_section}, one could consider that these 
orbits are chaotic. However, we will show that this conclusion is
false. Indeed, a careful identification of the \texttt{MEGNO} 
time evolution shows that the latter always approach slowly the limit
$2$ from above. The closer to the separatrice, the slower the 
convergence. More precisely, %none of the above simulated orbits
%%presents a \texttt{MEGNO} time-evolution around a linear divergence 
%line, leading to the conclusion that these orbits are actually
%unstable periodic orbits, and as a matter of fact also regular. 
orbits close to the separatrix integrated over long time span present a 
bounded \texttt{MEGNO} evolution. Hence they should be considered 
as non chaotic.

To clarify this point, we performed a similar study, but using a
significantly longer time-span, namely 300~years. The results are
shown in Figure~\ref{samplecs22}~(right panel). For the sake of
comparison, the color bars have been taken identical on both
plots. Let us observe that the maximum value reached by the
\texttt{MEGNO} is $4$ in the left panel and $2.5$ in the right one. 
In the 300~years simulation (Figure \ref{samplecs22}, right), the
\texttt{MEGNO} values, associated with orbits close to the
separatrices, turn out to be, on average, smaller than in
Figure~\ref{samplecs22}~(left panel), reaching almost the limit
$\overline{Y}(t) \rightarrow 2$, due to the longer time of
integration. Similarly, the dark zone in the neighborhood of the
stable equilibrium point, corresponding to \texttt{MEGNO} values close
to zero, is strongly shrunk, supporting the result that, in the
limit of infinitely large $t$, only the orbit originating from the
exact stable equilibrium point leads to $\overline{Y} = 0$, whereas the
neighboring trajectories converge slowly to $\overline{Y}(t) = 2$.

Let us note that the importance of the integration time has been
recently reported by \cite{barrio2007} in the framework of applications of the
\texttt{MEGNO} method. %, and it is here confirmed. 
And, we confirm that a too short time of integration can give 
wrong conclusions about the dynamical behavior. 
Moreover, the latter paper also underlines some 
spurious structures appearing in the maps of the variational chaos
indicators, explaining the presence of %some background patterns 
the sine wave of lower \texttt{MEGNO} with a bulge at the center
of Figures~\ref{samplecs22}, ``{\it suggesting that the same periodic
orbit is more or less regular depending on the initial conditions
choice}''. Actually, accordly to the latters authors and to our 
analysis, this conclusion is wrong because this spurious 
structure is related to numerical artifacts.

%%%%%%%%%%%%%%%%%%%%%%%%%%%%%%%%%%%%%%%%%%%%%%%%%%%%%%%%%%%%%%%%%%%%%%%%%%%
\section{High area-to-mass ratios analysis}
\label{high-area-to-mass-ratios-analysis} 
The study of the long-term stability of near-geosynchronous objects
has recently prompted an increasing interest of the scientific community. 
In the particular  
case of classical near-geosynchronous objects, the problem has been
solved by computing the \texttt{MEGNO} indicator for a family of
simulated geostationary, geosynchronous and super-geosynchronous
orbits. A classical near-geosynchronous object has a period close 
to one sidereal day and is subjected to the main gravitational 
effects of the Earth, including the 1:1 resonance, luni-solar 
perturbing effects, as well as solar radiation pressure 
associated to a small area-to-mass ratio ($A/m \ll
1~\mathrm{m^2/kg}$). According to \citet{breiter2005} and
\citet{wytrzyszczak2007}, the near-geostationary region presents
chaotic orbits only very close to the separatrices, due to the
irregular transits between the libration and the circulation
regimes. Regarding the super-geostationary orbits, all of them seem
to be entirely regular on the time scale of the investigations, that
is a few decades.\\

The aim of this section is to provide a more
extensive analysis of the dynamics of near-geosynchronous space 
debris with high area-to-mass ratios ($A/m \gg 1~\mathrm{m^2/kg}$), 
subjected to direct solar radiation pressure. Our main
objective is to study the effects of high area-to-mass ratios on the
stability of the principal periodic orbits and on the chaotic
components. This analysis is divided into three parts. First,
in section~\ref{sensitivity}, we focus our 
attention on the sensitivity to initial conditions; then, 
in section~\ref{extendedanalyses}, we report the results of dedicated 
numerical analyses 
which emphasize the importance of the area-to-mass ratio
value. Finally, in section~\ref{initialec}, we study the influence
of both the initial eccentricity and time at epoch.

Let us recall that for large area-to-mass ratios ($A/m \geq
10~\mathrm{m^2/kg}$), the solar radiation 
pressure may become the major perturbation, by far larger than the
dominant zonal gravity term $J_2$ \citep{valk07b}. In this particular 
case, the larger the area-to-mass ratio, the more affected the
dynamics of the near-geosynchronous space debris, leading to daily 
high-amplitude oscillations of the semi-major axis, yearly
oscillations of the eccentricity as well as long-term variations of 
the inclination.
\begin{figure}[!htbp] 
    \begin{tabular*}{\textwidth}{c}
      \includegraphics[width=\textwidth]{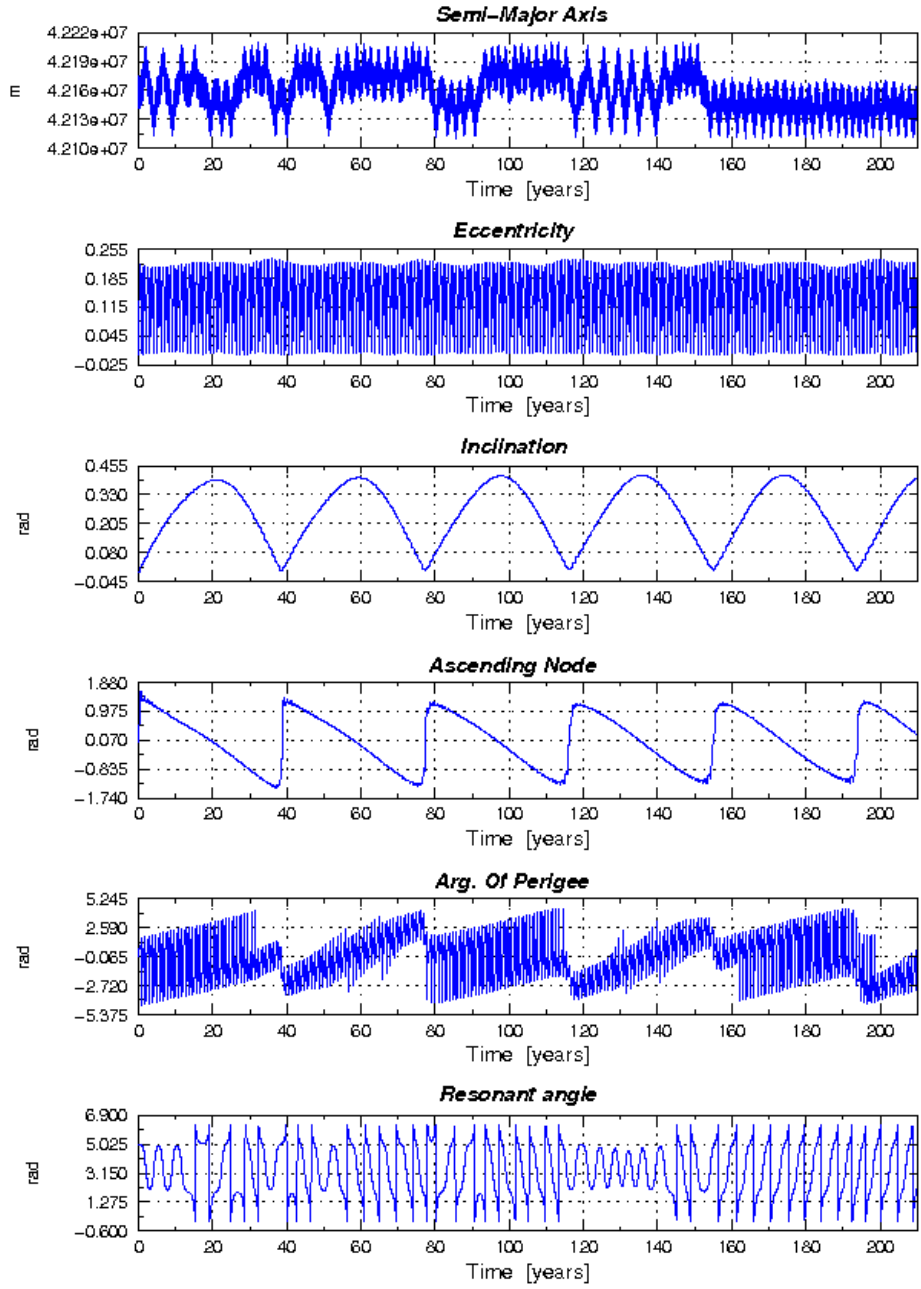} 
    \end{tabular*}
    \caption{\label{sensitivity_orbit} Time-evolution of high 
      area-to-mass ratio space debris. Orbital elements over
      $210$~years for a $A/m = 10~\mathrm{m^2/kg}$; initial conditions
      are: $a_0 = 42166.473$~km, $e_0 = 0.002, i_0 = 0.004$~rad, $\Omega_0 =
      \omega_0 = 0$~rad and $M_0 = 4.928$~rad. Time at epoch is 
      25~January~1991.}
\end{figure}
As an illustration, Figure~\ref{sensitivity_orbit} shows the orbital
elements histories of the first $210$~years of a 
geosynchronous high area-to-mass ratio space debris ($A/m =
10~\mathrm{m^2/kg}$). The yearly variation of the eccentricity reaches
$0.2$, which confirms the expected values predicted 
%(e.g. \citealt{anselmo05}, and \citealt{liouweaver05}). The
(e.g. Anselmo and Pardini, 2005, and Liou and Weaver, 2005). The 
inclination evolution presents a well known long-term variation whose
period is directly related to the area-to-mass ratio value. Regarding
the longitude of ascending node as well as the argument of perigee,
they both present a libration due to the chosen set of initial
conditions. For further details, we refer to \citet{valk07b} as well
as \citet{chao06}, where a full description of the long-term motion of
high area-to-mass ratios space debris is given.
%--------------------------------------------------------------------- 
%---------------------------------------------------------------------
%--------------------------------------------------------------------- 
%---------------------------------------------------------------------
\subsection{Sensitivity to initial conditions}\label{sensitivity} 
\begin{figure}[!htp]
  \begin{center}
      \includegraphics[width=\textwidth]{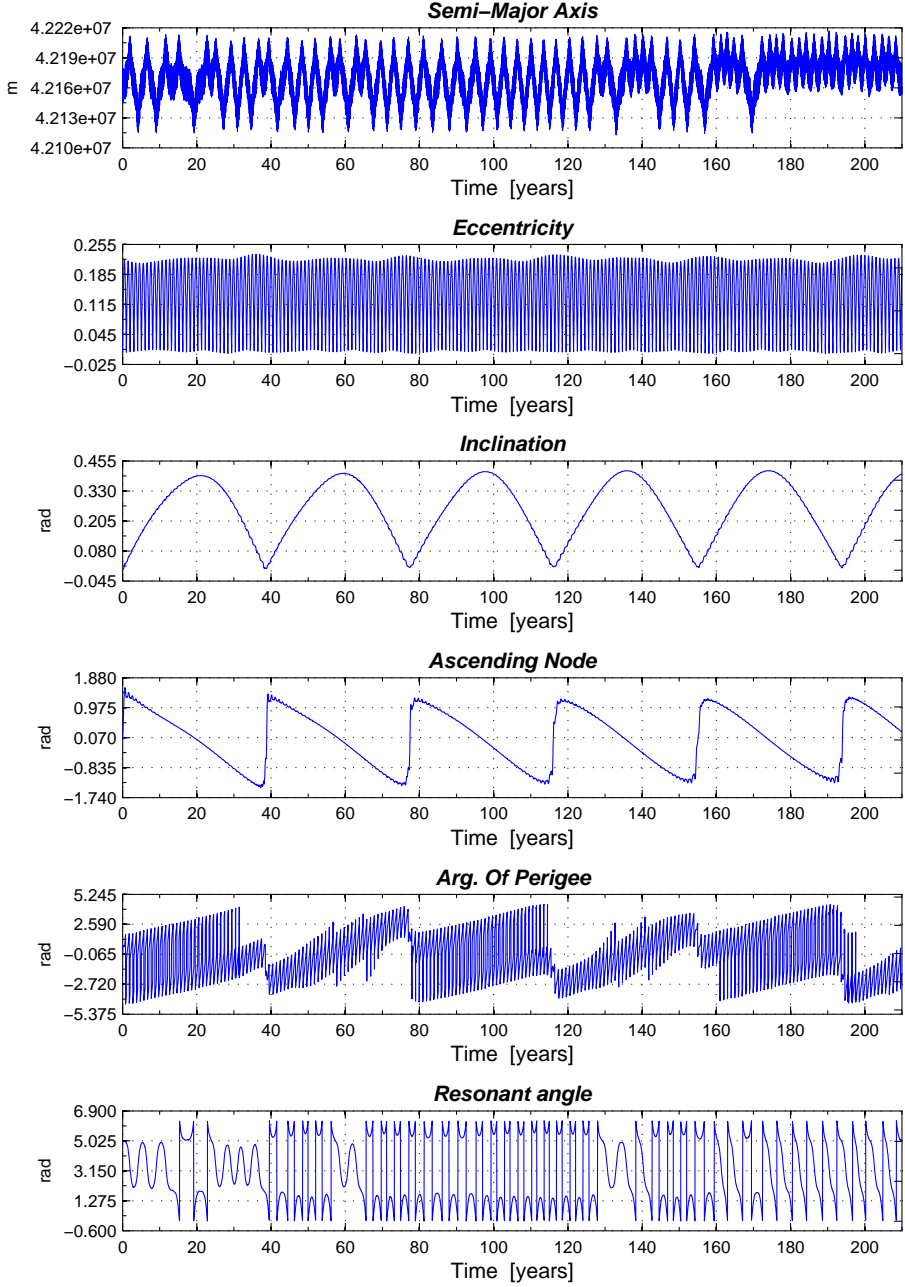}
  \end{center}
    \caption{\label{sensitivity_orbit2}  Time-evolution of high 
      area-to-mass ratio space debris. Orbital elements over
      $210$~years for a $A/m = 10~\mathrm{m^2/kg}$; the initial conditions 
      are the same of Figure~\ref{sensitivity_orbit} but differ in the 
      10th digits in mean longitude.}
\end{figure}

To start with, we follow the evolution of two high area-to-mass
ratio space debris ($A/m=10~\mathrm{m^2/kg}$) defined by two sets of
very close initial conditions, differing only in the 10th digits in mean
longitude. Figure~\ref{sensitivity_orbit} shows the first one and 
Figure~\ref{sensitivity_orbit2} shows the second near orbit. We observe 
that the most difference (in the behavior) take place in the semi-major 
axis and resonant angle panels. We notice that there are some differences 
in the dynamics of the semi-major axis already after $20$ years and at 
the end of the integration. This is the same for resonant angle, 
confirming the hypothesis that the sensitivity to initial conditions 
is especially relevant for the semi-major axis and resonant angle 
whereas the difference between the other orbital elements remains small.  
We first focus our attention on the time evolution of the semi-major 
axis and resonant angle. 
As a complement to Figure~\ref{sensitivity_orbit}, we numerically
computed two orbits for two space debris with different area-to-mass
ratios, $A/m= 1~\mathrm{m^2/kg}$ and $A/m=10~\mathrm{m^2/kg}$, whose
initial conditions have been chosen near the separatrices, to emphasize
their chaotic behaviors.\\

\begin{figure}[!h]
    \begin{tabular*}{\textwidth}{c}
      \includegraphics[width=0.94\textwidth]{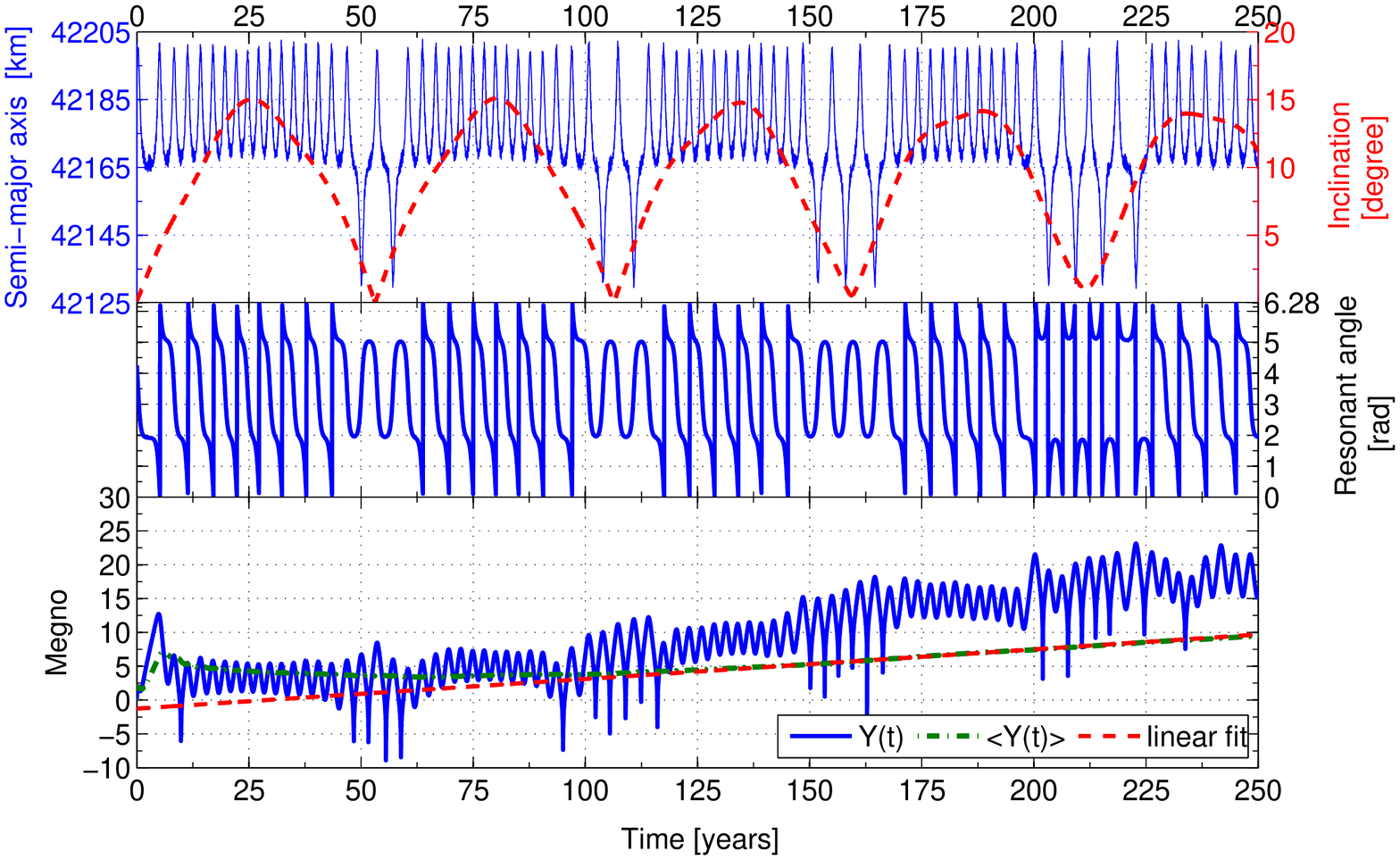}
      \vspace{-0.3cm}
      \\
      \vspace{-0.4cm}
      \includegraphics[width=0.94\textwidth]{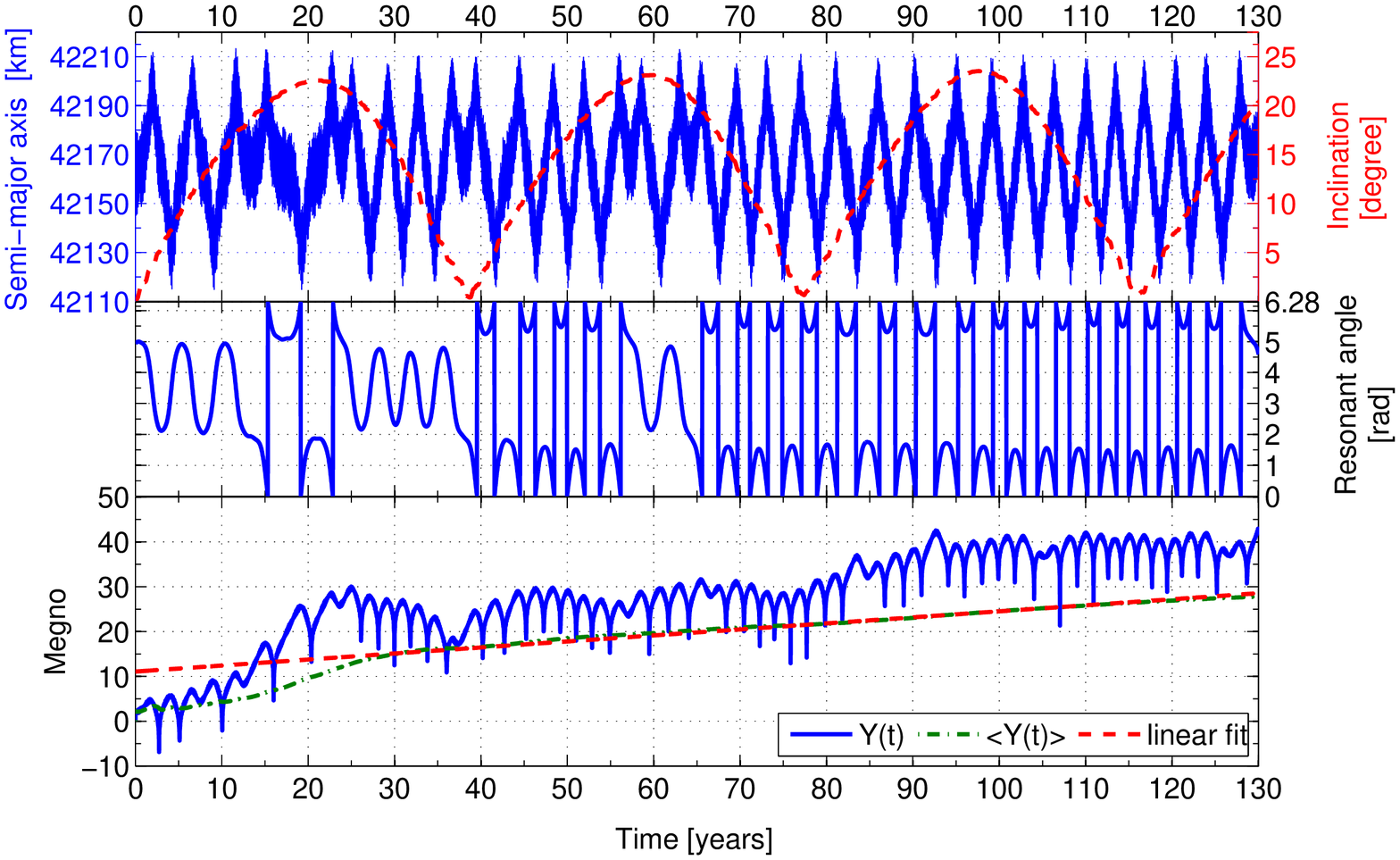}
    \end{tabular*}
    \caption{\label{transit_regression_1_10}For each graph, we show
    the orbital evolution of the semi-major axes $a$ (solid line)
    superimposed with the evolution of the inclinations (dashed line)
    [top panels]. The time evolution of the resonant angles [middle
    panels] and the time evolution of the \texttt{MEGNO} indicator
    ($Y$ and $\overline{Y} =<Y(t)>$), as well as the corresponding linear fit
    $\overline{Y}(t) \simeq a_\star \, t+d$ [bottom panels]. The
    area-to-mass ratios are $A/m=1~\mathrm{m^2/kg}$ in the upper panel and
    $A/m= 10~\mathrm{m^2/kg}$ in the lower one. The initial conditions are
    chosen 
    near the separatrices. The computed linear regression coefficients
    are given by $a_\star=0.043$ (for $A/m=1~\mathrm{m^2/kg}$) and
    $a_\star=0.134$ (for $A/m=10~\mathrm{m^2/kg}$).}
\end{figure}

Figure~\ref{transit_regression_1_10} shows a
blow-up of the evolution of the semi-major axis (top panels) and 
resonant angle (middle panels) over the time span of
$250$~years. It is clear that the semi-major axis presents some
irregular components over its evolution, related to some transitions
between different regimes of motion, clearly identifiable in the
resonant angle plots. In addition, we also computed the corresponding
\texttt{MEGNO} time evolution. The bottom panel in each graph shows
the time evolution of the \texttt{MEGNO} indicator as well as its
corresponding mean value. 
First, we see that the time evolution of $\overline{Y}(t)$ presents a
quasi-linear growth almost since the beginning of the integration
process, leading to the conclusion that these orbits are clearly
chaotic over that time scale.\\
\newpage
Therefore, we also computed the linear 
fit $\overline{Y}(t) \simeq a_\star \, t+d$ %in both cases 
in order to evaluate the Lyapunov time $T_\lambda$. $T_\lambda$ is the 
inverse of the LCN ($\lambda$) calculated by the linear regression 
coefficients $a_\star = \lambda /2$. 
Let us remark that to avoid the initial transient state, the least
square fits were performed on the last 85\% of the time interval. This
latter analysis brings to the fore the fact that larger area-to-mass
ratios lead to smaller Lyapunov times, i.e. larger Lyapunov
characteristic numbers. Indeed, for $A/m = 1~\mathrm{m^2/kg}$, the
Lyapunov time turns out to be on the order of $11$~years, whereas it
reaches the value $T_\lambda \simeq 3.7$~years for $A/m =
10~\mathrm{m^2/kg}$.
\vspace{-0.2cm}
\newpage

Second, let us also remark that the behavior of
the \texttt{MEGNO} indicator is of particular interest in 
these cases. A careful analysis of $Y(t)$ underlines some
irregular patterns directly related to the evolution of $\sigma$, in
particular when the orbits seem to transit across the
separatrices. Finally, we can also highlight the fact that the sudden
changes between libration and circulation regimes occur mainly when
the inclination changes its sign of variation, especially at the
maximum value for $A/m >> 1~\mathrm{m^2/kg}$ and at the minimum for
$A/m \leq 1~\mathrm{m^2/kg}$ (Figure~\ref{transit_regression_1_10},
top panels, dashed line), with an empirical long-term periodicity of
$T_\Omega$, that is the long-term periodicity of the longitude of the
ascending node, which is all the more smaller when $A/m$ is large
\citep{valk07b}.

\subsection{Extended numerical analyses}\label{extendedanalyses}
\begin{figure}[!htb] 
  \begin{center}
%  \setstretch{.7} 
    \begin{tabular}{ll}
      \includegraphics[width=8cm]{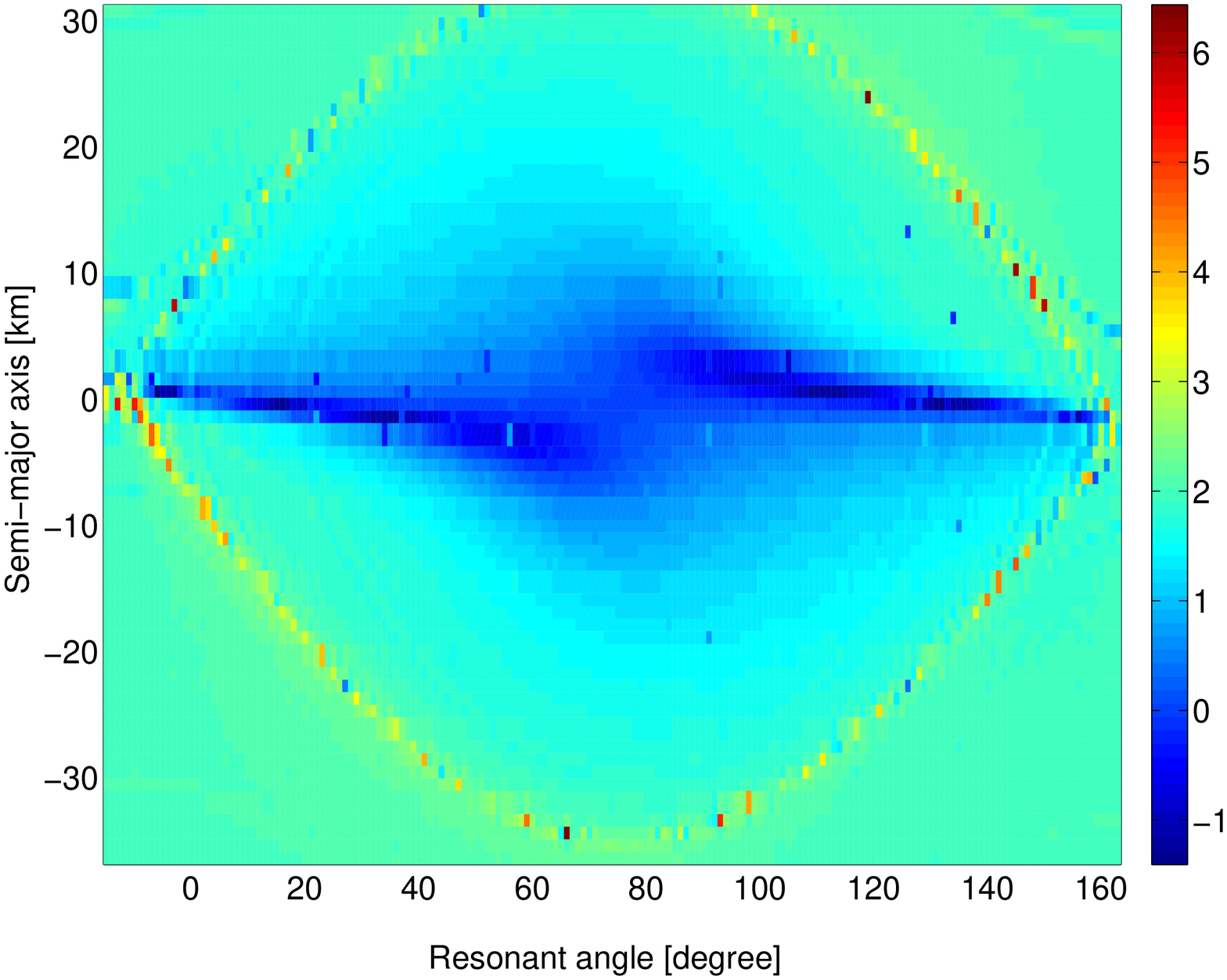}&
      \includegraphics[width=8cm]{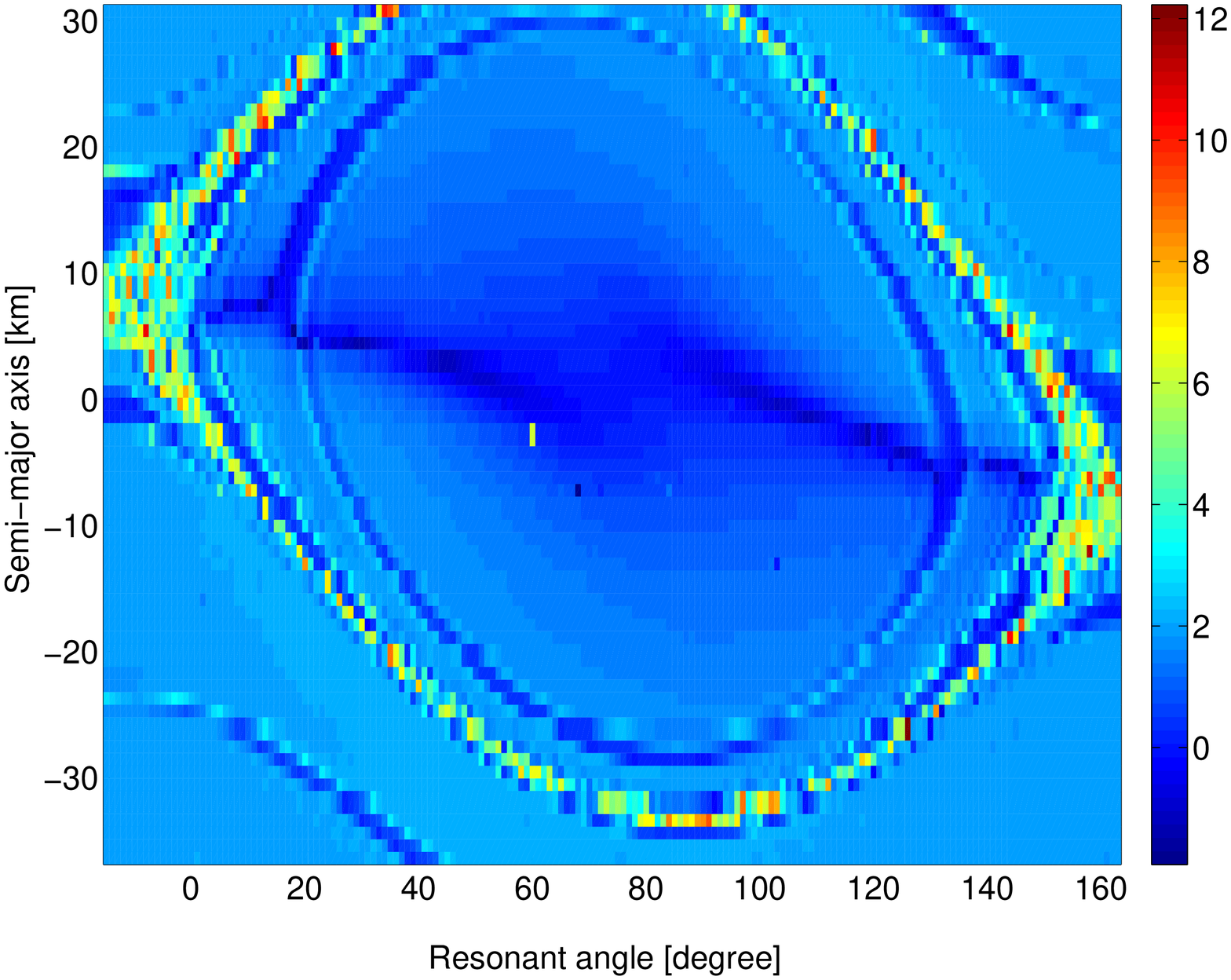}\\ 
      \includegraphics[width=8cm]{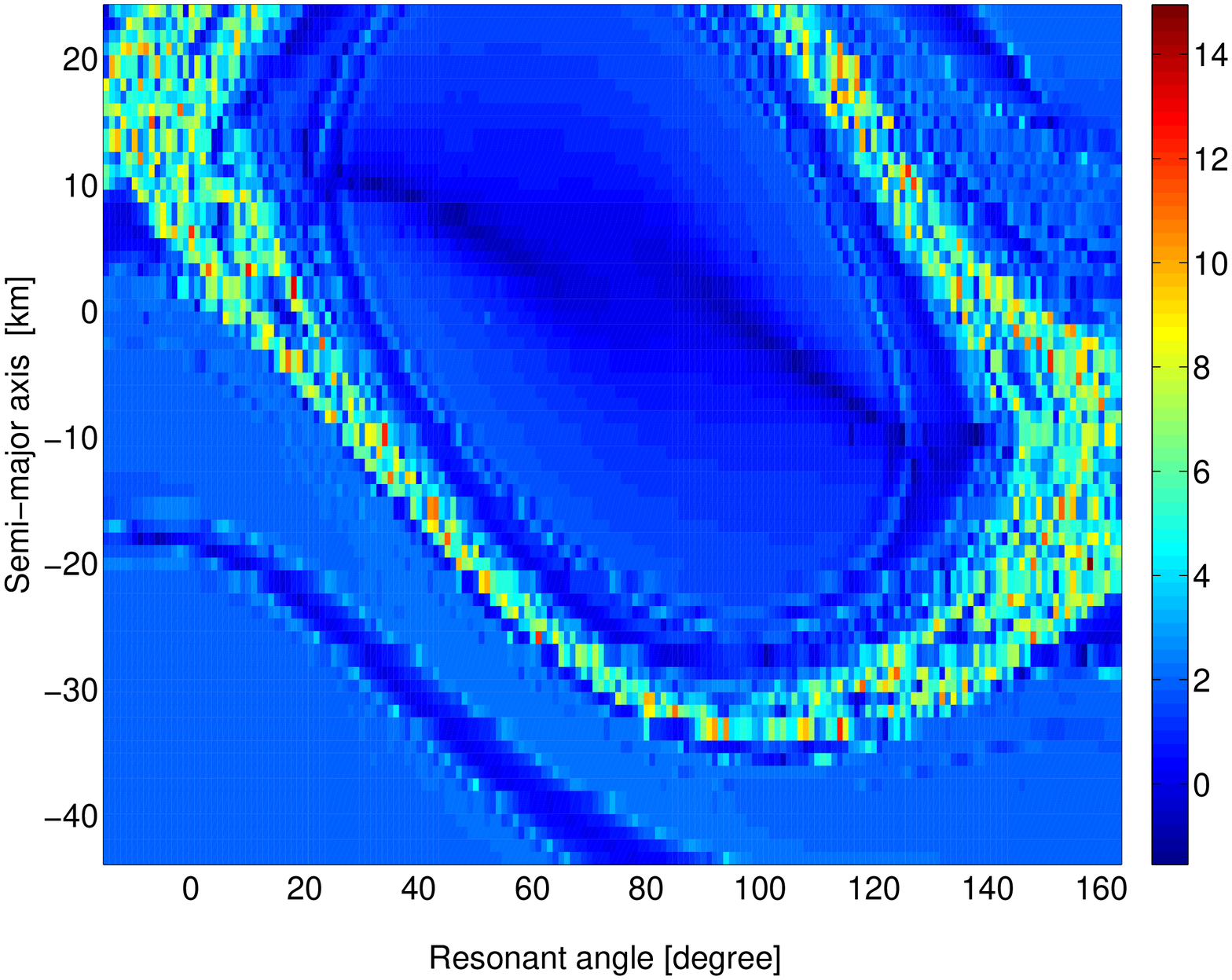}& 
      \includegraphics[width=8cm]{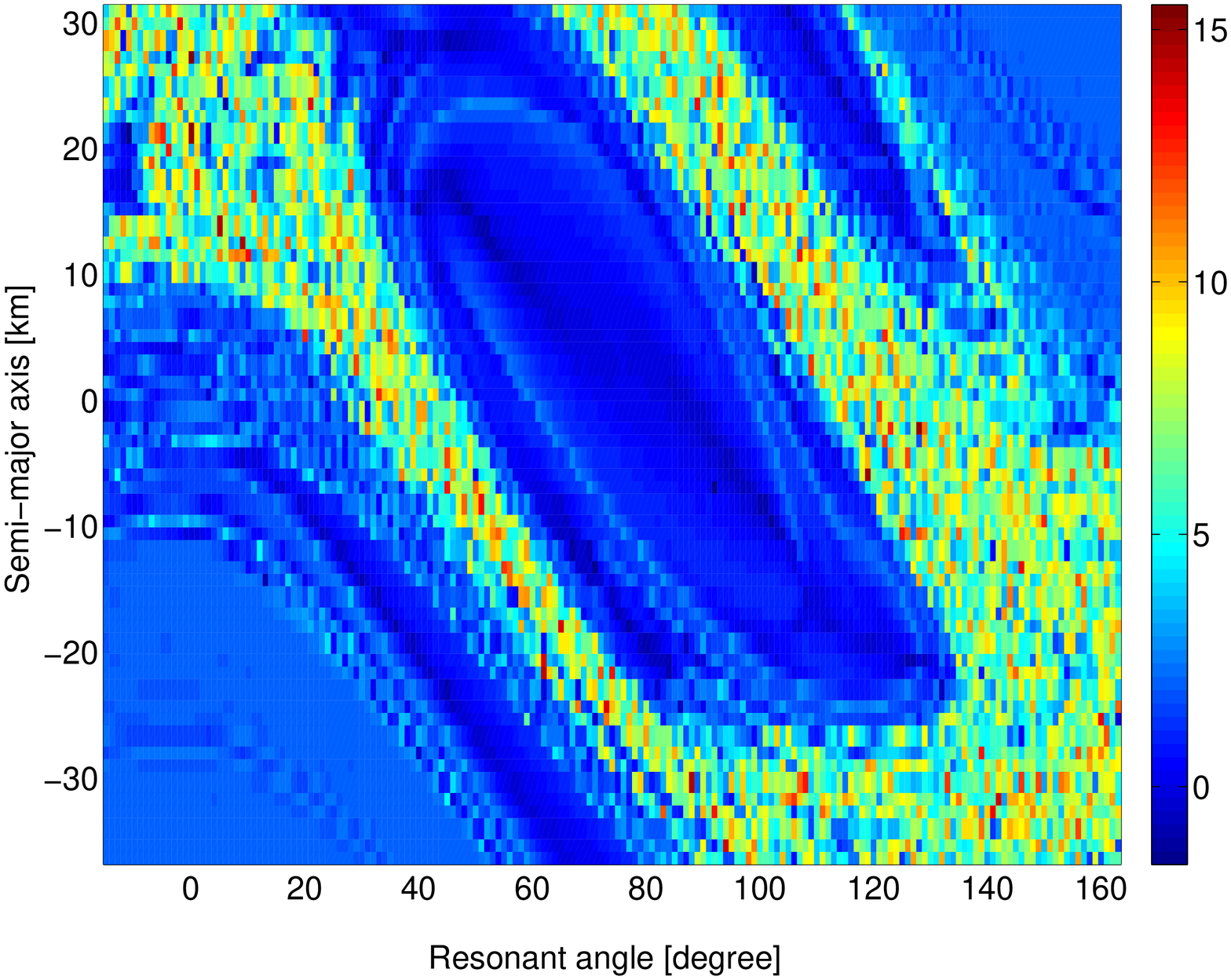}
    \end{tabular} 
    \caption{\label{sample_various_asm}The \texttt{MEGNO} computed as
    a function of initial mean longitude $\lambda_0$ and initial
    (osculating) semi-major axis $a_0$. The equations of motion
    include the central body attraction, the second degree and order
    harmonics $J_2,C_{22}$ and $S_{22}$, the luni-solar interaction as
    well as the perturbing effects of the solar radiation
    pressure. The mean longitude grid is $1^{\circ}$ and the
    semi-major axis grid is $1$~km, spanning the $42164 \pm 35$~km
    range. The initial conditions are $e_0 = 0.002, i_0 = 0.004$~rad and 
    $\Omega_0 = \omega_0 = 0$~rad. The integration time is 30~years from
    epoch fixed at 25~January~1991. The patterns have been obtained
    using four different area-to-mass ratios, $A/m = 1, 5, 10,
    20~\mathrm{m^2/kg}$, represented, respectively in the top left, 
    top right, bottom left and bottom right panel.}
  \end{center}
\end{figure}

We considered a set of $12\,600$ simulated orbits with various initial
semi-major axes and mean longitudes. We took into account the following 
perturbing effects: second degree and order harmonics ($J_2,C_{22}$ and 
$S_{22}$), the luni-solar interaction as well as the perturbing effects 
of the solar radiation pressure with four values of the area-to-mass 
ratio ($A/m=1,5,10,20~\mathrm{m^2/kg}$). The results are 
reported in Figure~\ref{sample_various_asm}.\\

In the case with $A/m =
1~\mathrm{m^2/kg}$ (top left panel) we recognize the same pendulum-like
pattern as in 
Figure~\ref{samplecs22}. Considering the same integration time
(30~years), we notice that the \texttt{MEGNO} values tend to be 
slightly larger than in Figure~\ref{samplecs22}~(left). Moreover, some
irregularly distributed \texttt{MEGNO} values are clearly visible 
close to the two saddle unstable stationary points.
%--------------------------------------------------------------------------------
These results completely agree with those presented by
\citet{breiter2005}, where the solar radiation pressure was taken into
account, but only for very small area-to-mass ratios (typically 
$0.005~\mathrm{m^2/kg}$). Indeed, our latter analysis shows that in
addition to the luni-solar perturbations, solar radiation pressure 
with small to moderate area-to-mass ratios, that is $0 \le A/m \le
1~\mathrm{m^2/kg}$, do not change considerably the phase space 
pattern.
 
On the other hand, the remaining panels of Figure~\ref{sample_various_asm} show
that the phase portrait becomes significantly more intricate with
increasing area-to-mass ratios. Indeed, the width of the stochastic
zone in the neighbourhood of the separatrices becomes relevant, with a
large displacement of the separatrices on the phase plane. The larger
chaotic region can readily be explained by the osculating motion of 
the separatrices due to the before-mentioned daily variations of the
semi-major axis with respect to some mean value as well as by the 
increasing amplitudes of the eccentricities. These variations lead
inevitably to transits through both the regions separating libration 
and circulation motion for orbits initially close to the separatrices.
 
\begin{figure}[!ht]
  \begin{center} 
    \includegraphics[width=\textwidth]{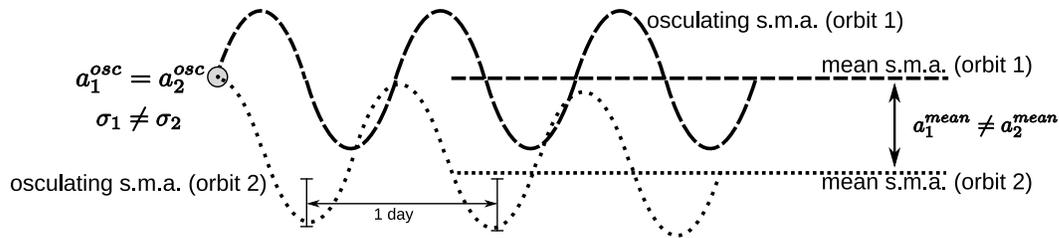}
  \end{center} 
  \caption[]{\label{meanosc}Cartoon to illustrate the difference
    between mean and osculating initial conditions with respect to the
    semi-major axis (s.m.a.) evolution. For the sake of simplicity, the
    mean semi-major axis does not present any long-term variation, whereas 
    the osculating semi-major axis presents daily oscillations related to
    direct solar radiation pressure (the implicit underlying model is
    radiation pressure only). It is clear that even if the osculating initial
    conditions $a_1^{osc}$ and $a_2^{osc}$ are 
    identical, the corresponding mean initial conditions $a_1^{mean}$ and
    $a_2^{mean}$ can be significantly different, due to different initial
    mean longitudes (or similarly different initial resonant angle values).}
\end{figure}

Moreover, it is also clear that the usual double pendulum-like phase space
shows a tendency to be distorted with an apparent displacement of the
unstable equilibrium points, whereas the stable equilibrium points
remain almost fixed. This last result is however quite awkward insofar
as there is no physical interpretation to this phenomenon. Indeed, 
direct solar radiation pressure does not depend explicitly on the (mean) 
resonant angle with respect to the long-term investigations 
after averaging over short perioding terms. Therefore, it can
not induce a displacement of the equilibrium points in the phase
space. Actually, a clever explanation can be found regarding the
way the sampling is considered in the elaboration of the
graphics. More specifically, it is worth noting that, at first, the
sampling is carried out with respect to osculating initial
conditions.\\

Second, within the framework of mean motion theory, it is
well-known that, due to the short-period oscillations, the mean and
the osculating initial conditions can not be considered to be
equal. In other words, for the same fixed value of the initial 
osculating semi-major axis and for various initial mean longitude, we 
obtain different values for the mean semi-major axis;
%In other words, when considering an horizontal line in the
%initial conditions sampling, even though it corresponds to a fixed
%value of the initial osculating semi-major axis, it is actually
%related to a various set of mean initial semi-major axes 
as explained with Figure~\ref{meanosc}. Actually, the 
different initial mean longitudes induce a phase difference in the
corresponding evolution of the semi-major axis, leading to different
mean initial semi-major axes. Let us remark that the maximum
difference between both the mean semi-major axes is directly related
to the order of magnitude of the short-period variations, and, as a
consequence, is also directly related to the area-to-mas ratio.
\begin{figure}[!t]
  \begin{center}
    \includegraphics[width=\textwidth]{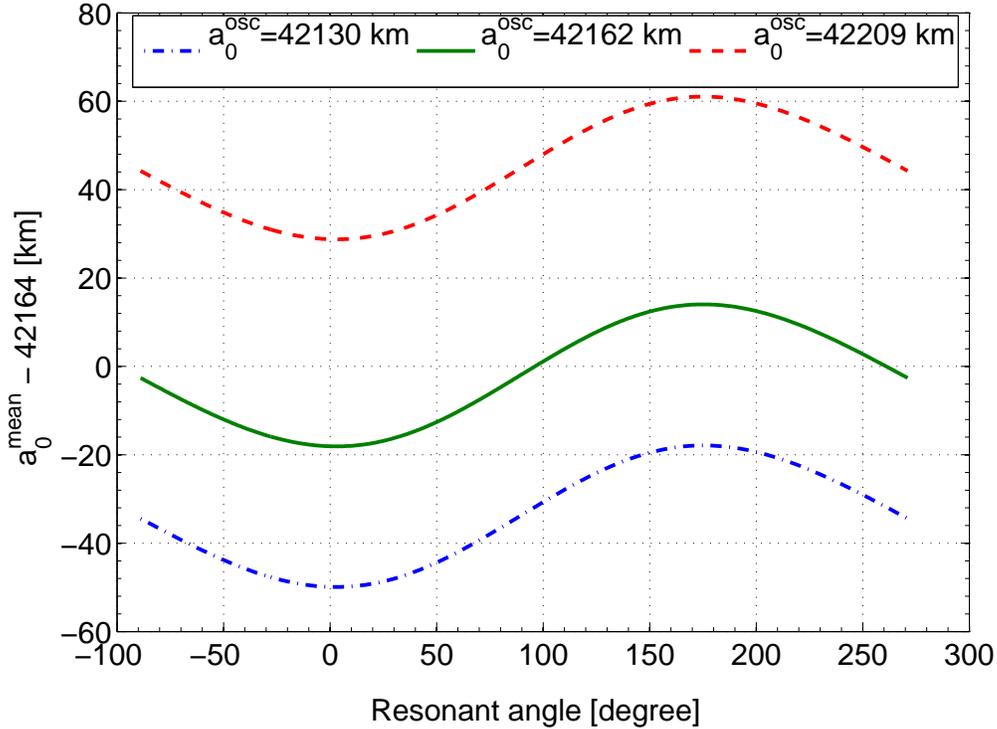}
  \end{center}
  \caption[]{\label{meanosc2}Relation between the mean semi-major axis
    and the resonant angle for various values of the osculating semi-major
    axis. The first osculating semi-major axis is taken above the
    libration region, the second is related to an osculating semi-major
    axis sampling which crosses the libration region and, finally, the
    third sampling is taken below this region.}
\end{figure}\\

More rigorously, the difference between osculating and mean initial
conditions is a well-defined transformation, depending on the
generating function used within the averaging process allowing to
change from mean to osculating dynamics. For further details
concerning this explicit transformation, we refer to the Lie algorithm
discussed in \citet{deprit69} and \citet{henrard70}. However, because
we bound our analysis mainly to numerical simulations, we cannot
access such generating function; we can nevertheless overcome this
problem by numerically computing, for each semi-major axis osculating
initial condition, the related mean initial semi-major axis, by
considering the average over a short time span of 10 days. As an
illustration, in Figure~\ref{meanosc2}, we give the relation between
the mean semi-major axis and the resonant angle for various values of
the osculating semi-major axis ($A/m = 10~\mathrm{m^2/kg}$). The first
difference is related to a semi-major axis sampling taken above the
libration region, the second is related to a semi-major axis sampling
which crosses the libration region and finally, the third sampling is
taken below this region. In conclusion, we clearly see that the order
of magnitude of the differences is, as previously mentioned, the order
of the amplitudes of the daily variations observed in the semi-major
axis dynamics. Let us note that in the latter case, i.e. $A/m =
10~\mathrm{m^2/kg}$, the differences reach at most $27$~km, which
correspond exactly to the difference between the stable and unstable
equilibrium points, as shown in
Figure~\ref{sample_various_asm}~(bottom, left).

\begin{figure}[!t]
  \begin{center}
    \begin{tabular}{cc}
      \includegraphics[width=.46\textwidth]{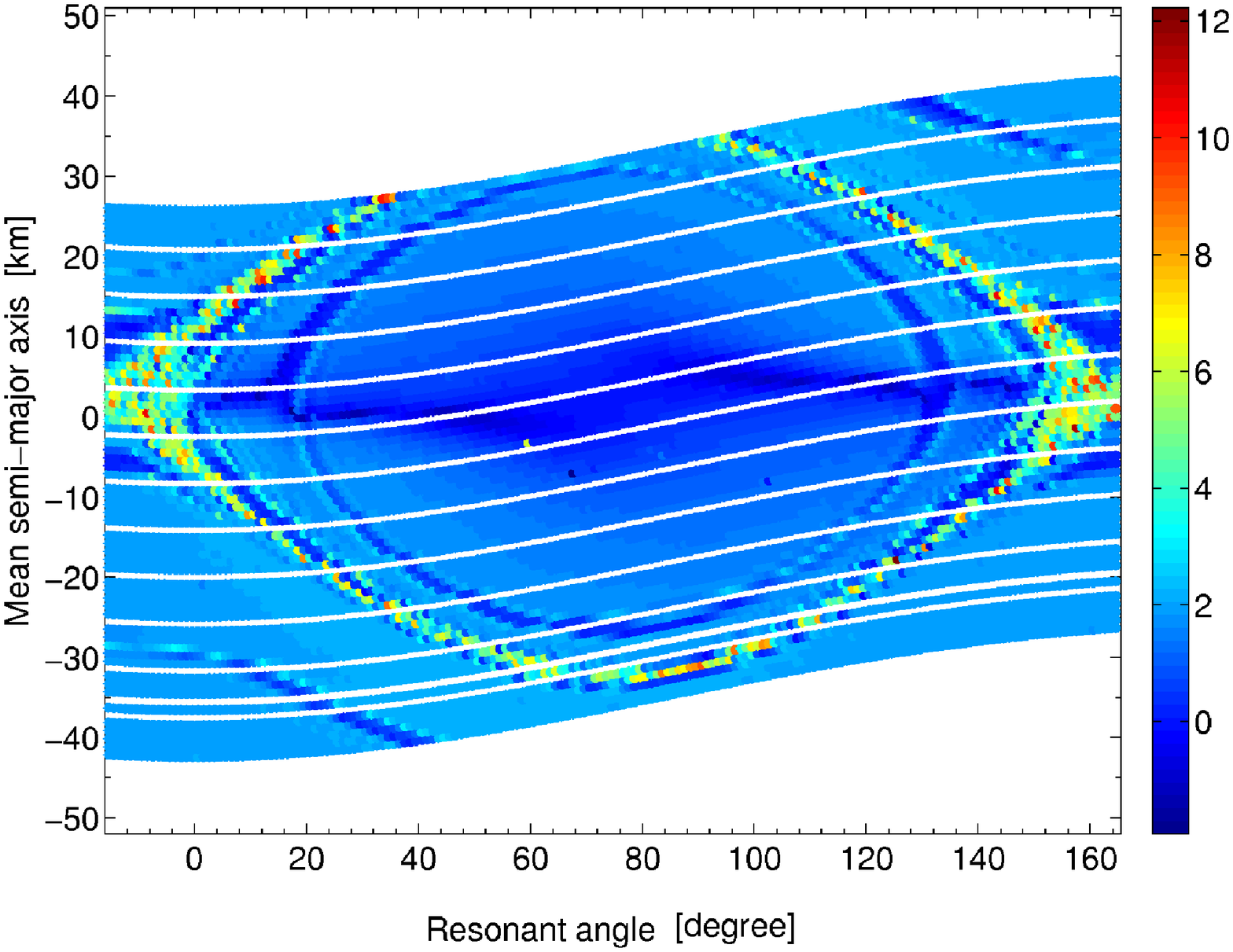}&
      \includegraphics[width=.46\textwidth]{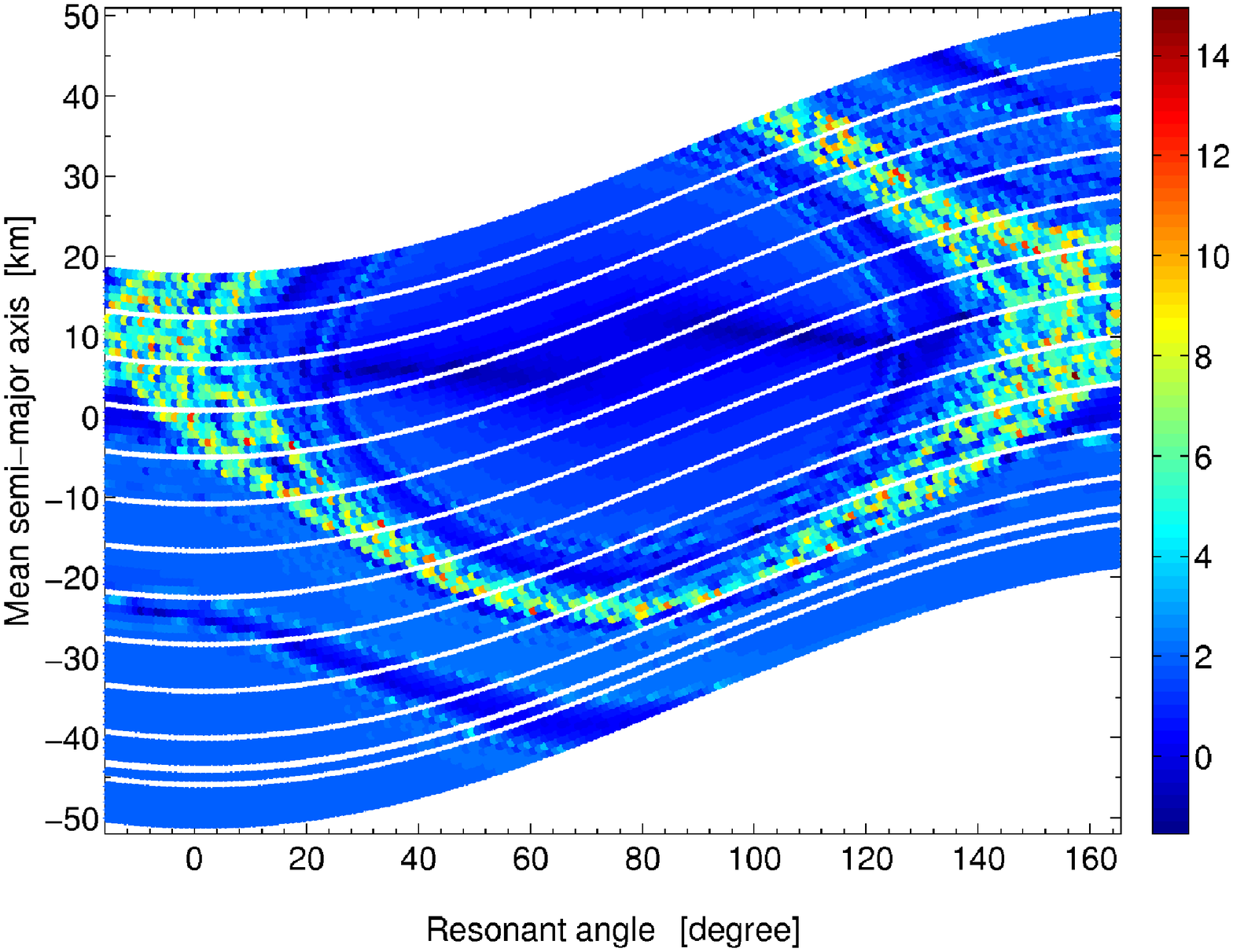}
    \end{tabular}
    \caption{\label{sample_various_asm_transf}The \texttt{MEGNO}
    computed as a function of initial mean longitudes $\lambda_0$ and
    initial mean semi-major axis $a_0$. The model is the same as in
    Figure~\ref{sample_various_asm}. The area-to-mass ratio is $A/m = 5$ 
    and $10~\mathrm{m^2/kg}$ for the left and for the right graph,
    respectively.}
  \end{center}
\end{figure}

We can thus apply numerically the transformation as a
post-treatment process, that is considering the \texttt{MEGNO} values 
not in the osculating initial conditions phase space, but in the mean
initial conditions phase space. For the sake of comparison with
Figure~\ref{sample_various_asm}, we show the results once such a
transformation has been applied (Figure~\ref{sample_various_asm_transf}): it
is  clear that now the vertical gaps between both the stable and unstable
equilibrium points are almost completely eliminated, hence these points
have almost the same mean semi-major axis, getting rid of the what we called
the ``{\it short-period artefact}''. 
The thin light waves crossing the Figure~\ref{sample_various_asm_transf} 
are due to gaps in the set of initial conditions and have no 
dynamical significance (also valid for the only one light wave 
crossing the Figure~\ref{secondary_resonances} in section 
\ref{secondary_resonances_section}). Let us also remark that, 
from now on, all the results will be shown in the mean 
initial conditions phase space. 
%%%%%%%%%%%%%%%%%%%%%%%%%%%%%%%%%%%%%%%%%%%%%%%%%%%%%%%%%%%%%%%%%%%%%%%%%%%%%%%%%%%%%%%%
 \subsection{Initial time at epoch and importance  of the mean
   eccentricity}\label{initialec} 
One should also recall that solar radiation pressure leads to a
theoretical equilibrium defined both in eccentricity $e_0$ and
longitude of perigee $\varpi_0$. The conditions leading to such an
equilibrium can be written as
\begin{equation*}
\left\lbrace
\begin{array}{ccl}
\displaystyle e_0 &=& \displaystyle \frac{3}{2}\,C_r\,P_r\,\frac{A}{m}\,\frac{1}{n \,a\,n_\odot} \,\cos^2 \frac{\epsilon}{2} \simeq 0.01 \, C_r\,\frac{A}{m}\,,\\
\displaystyle \varpi_0 &=& \displaystyle \lambda_\odot(0)\,.
\end{array}
\right.
\end{equation*}
where $n$ and $n_\odot$ are the angular motions of both the space
debris and the Sun respectively, $\epsilon$ is the obliquity of the
Earth with respect to the ecliptic and $\lambda_\odot(0)$ the initial
ecliptic longitude of the Sun. If these conditions are fulfilled, it 
%has been shown (\citealt{chao06}, and later \citealt{valk07b}), that the
has been shown (Chao, 2006, and later Valk et al., 2008), that the 
eccentricity vector $(e\, \cos \varpi, e\, \sin \varpi)$ remains
constant, leading to a fixed value of both the eccentricity and longitude
of perigee.\\

\begin{figure}[ht]
  \includegraphics[width=\textwidth]{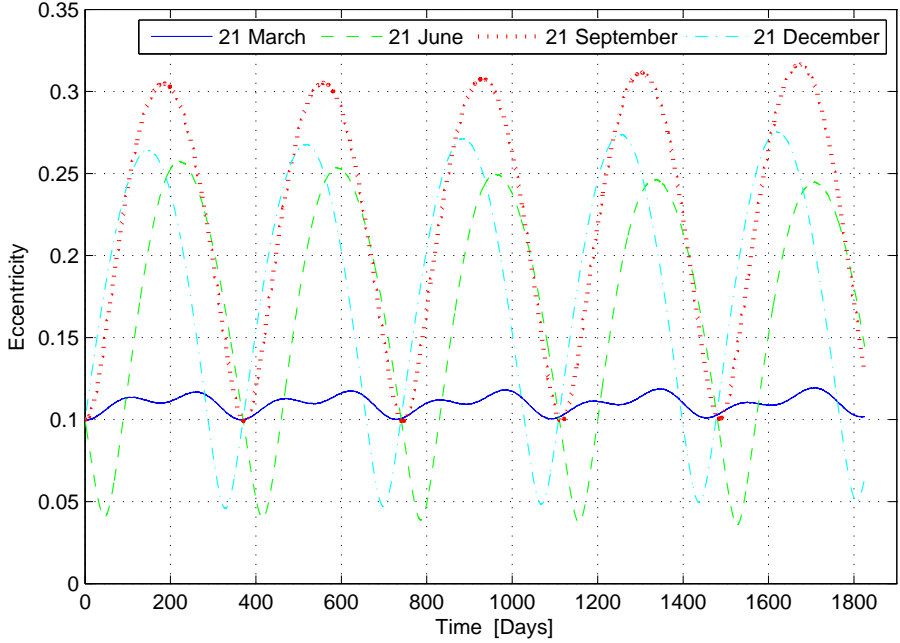}
  \caption{\label{eccentricity-equilibrium} Mid-term variations 
    (yearly oscillations) of the
    eccentricity for fixed initial conditions $a_0 = 42164$~km, $e_0 =
    0.1$, $i_0 = 0$~rad, $\Omega_0 = \omega_0 = \lambda_0 = 0$~rad and 
    $A/m = 10~\mathrm{m^2/kg}$. Various initial times at
    epoch $t_0$, namely different initial ecliptic longitudes of the Sun
    $\lambda_{\odot}(0)$, were used for the numerical propagations. 
    Only $J_2,C_{22},S_{22}$ and direct solar radiation pressure taken into 
    account.}
\end{figure}

As an illustration, Figure~\ref{eccentricity-equilibrium}
shows the mid-term variations of the eccentricity for a fixed value of
the area-to-mas ratio ($A/m = 10~\mathrm{m^2/kg})$ and fixed initial
conditions, namely, $a_0 = 42164$~km, $e_0 = 0.1$, $i_0 = 0$~rad, 
$\Omega_0 = \omega_0 = \lambda_0 = 0$~rad. It is clear that, apart
from a phase difference, the amplitudes of the variations of the
eccentricities are qualitatively the same, except when adopting an
initial time at epoch equal to 21 March. In this latter case, the
eccentricity remains almost constant, as expected by the theory.

\begin{figure}[ht]
  \begin{center}
    \begin{tabular}{ll}
      \includegraphics[width=8cm,height=6cm]{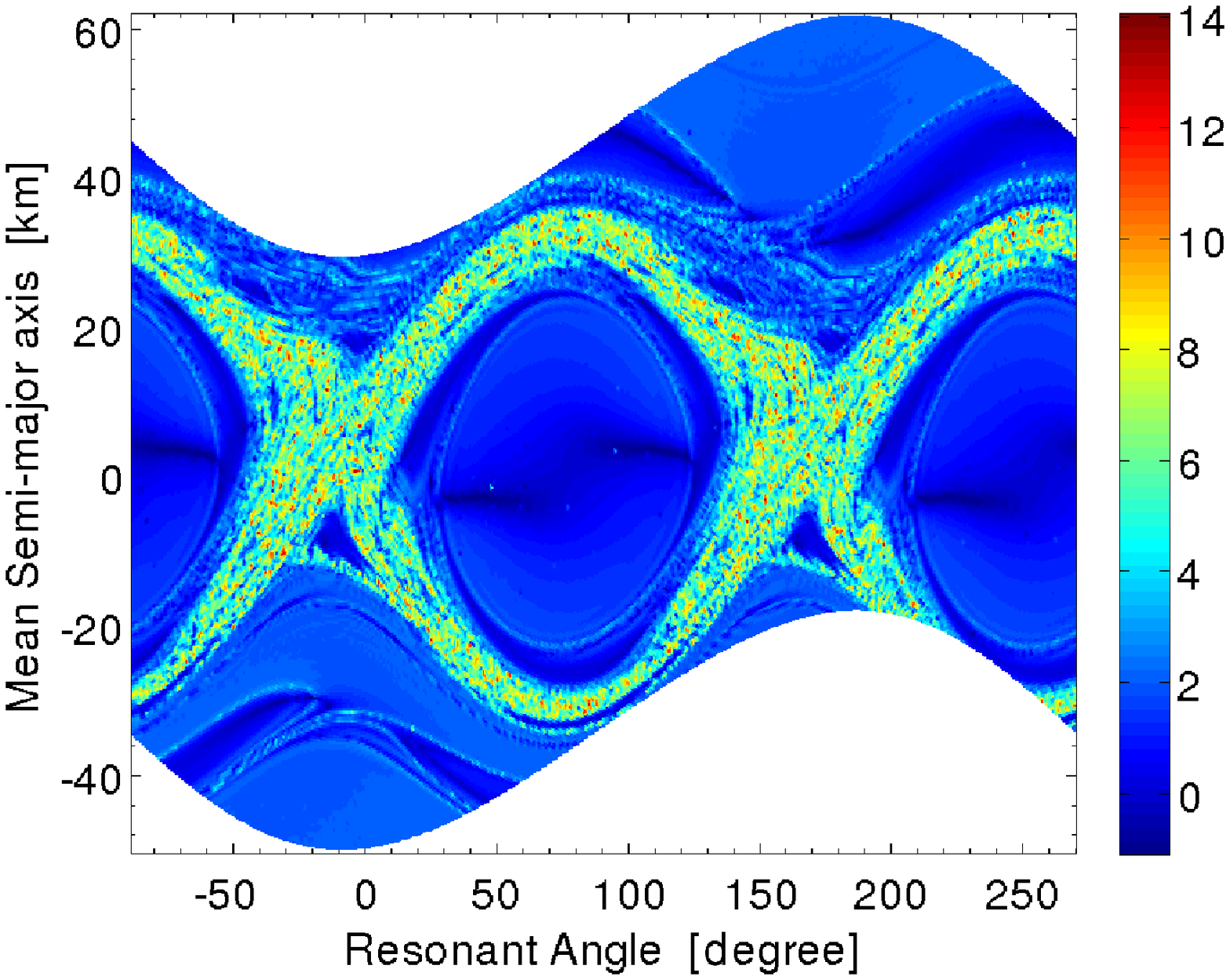}&
      \includegraphics[width=8cm,height=6cm]{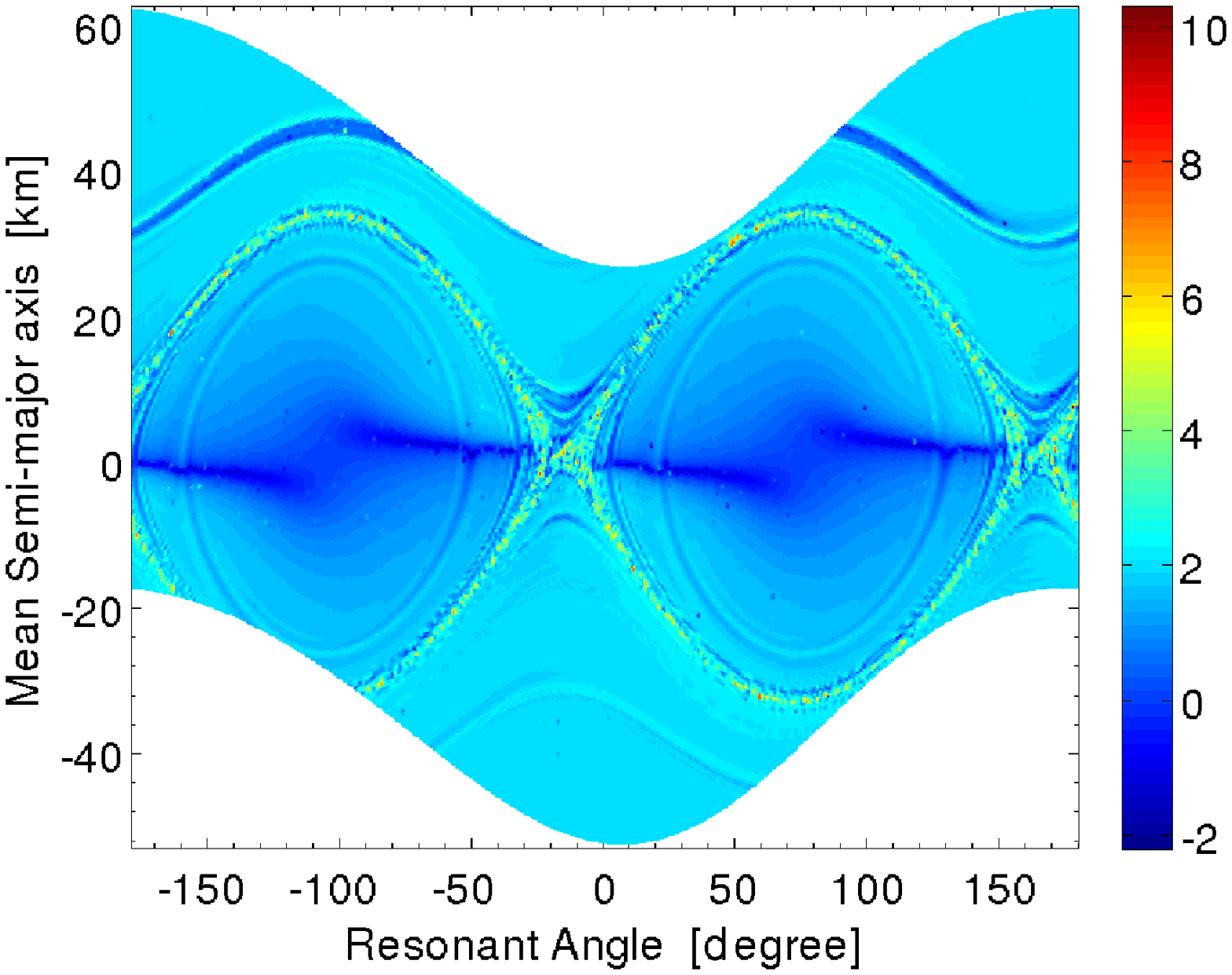}
    \end{tabular}
    \caption{\label{sample_epoch} \texttt{MEGNO} computed as a
    function of initial mean longitude $\lambda_0$ and semi-major
    axis $a_0$. The equations of motion include the central body
    attraction, the second degree and order harmonics $J_2,C_{22}$ and
    $S_{22}$, the luni-solar interaction as well as the perturbing
    effects of solar radiation pressure. The mean longitude grid
    is $1^{\circ}$ and the semi-major axis grid is $500$~m spanning the
    $42164 \pm 35$~km range. The initial conditions are
    $e_0 = 0.1, i_0 = 0.004$~rad, $\Omega_0 = \omega_0 = 0$~rad 
    with $A/m = 10~\mathrm{m^2/kg}$. The
    patterns have been obtained using two different initial times at
    epoch, namely 21~December~2000 (left) and 21~March~2000 (right),
    respectively.}
  \end{center}
%--------------------------------------------------------------------------
\end{figure}

\begin{figure}[!htbp]
  \begin{center}
    \begin{tabular}{ll}
      \includegraphics[width=8cm,height=6cm]{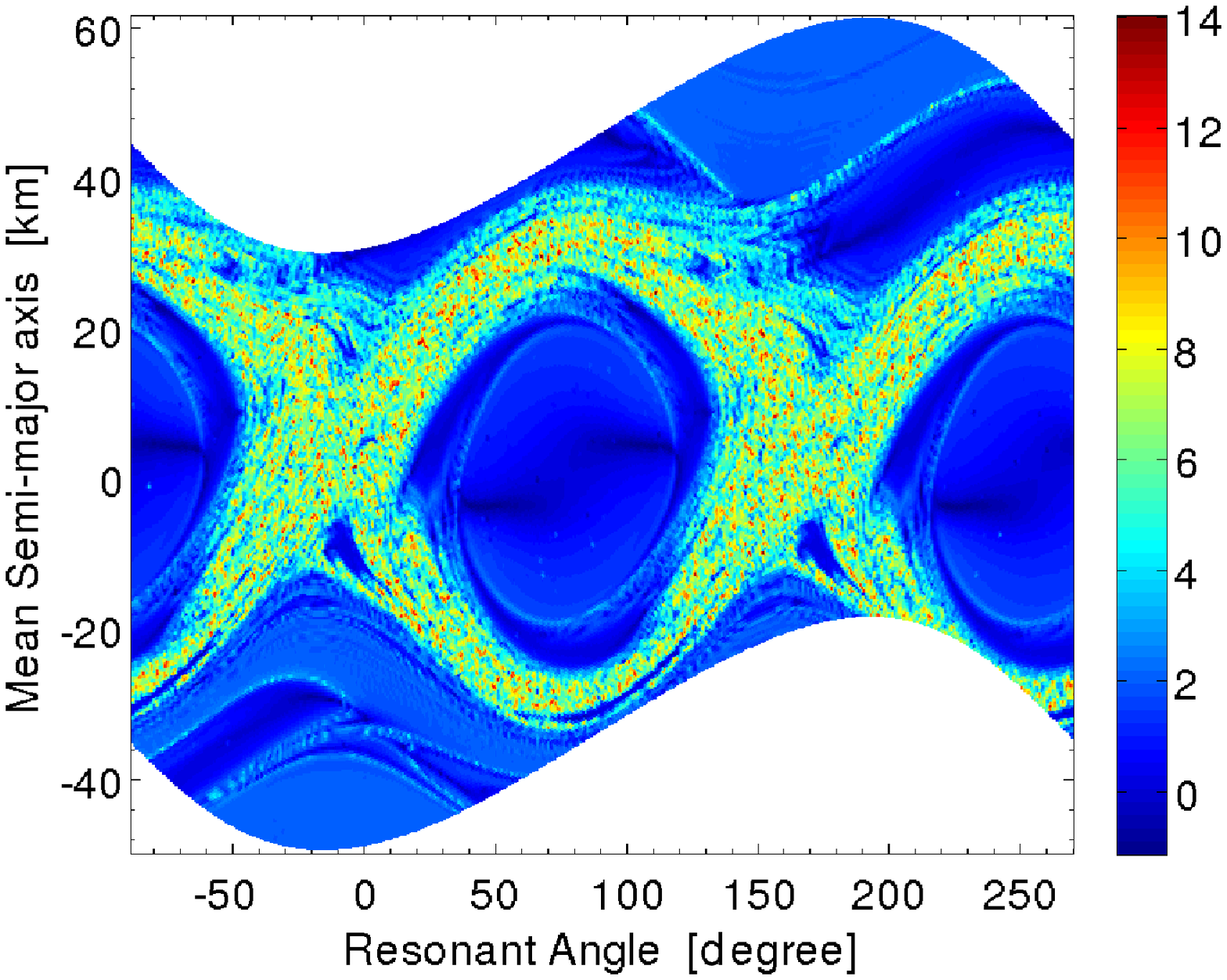}&
      \includegraphics[width=8cm,height=6cm]{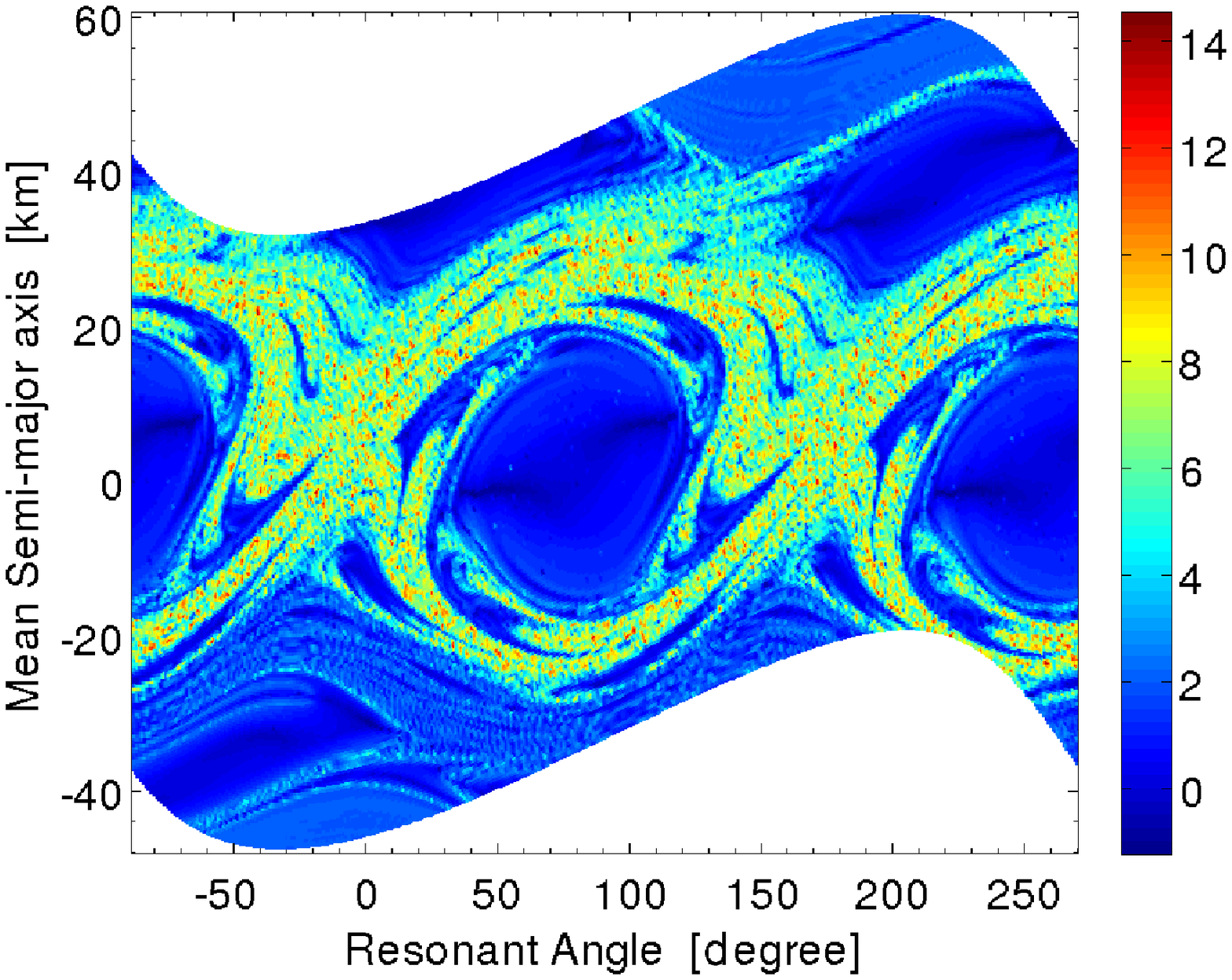}
    \end{tabular}
    \caption{\label{higher_eccentricities} \texttt{MEGNO} computed
    as a function of initial mean longitude $\lambda_0$ and
    semi-major axis $a_0$. The equations of motion include the central
    body attraction, the second degree and order harmonics
    $J_2,C_{22}$ and $S_{22}$, the luni-solar interaction as well as
    the perturbing effects of solar radiation pressure. The mean
    longitude grid is $1^{\circ}$ and the semi-major axis grid is
    $500$~m spanning the $42164 \pm 35$~km range. The initial
    conditions are $i_0 = 0.004$~rad, $\Omega_0 = \omega_0 = 0$~rad with 
    $A/m = 10~\mathrm{m^2/kg}$. Time at epoch is
    21~December~2000. The patterns have been obtained using two
    initial eccentricities, $e_0 = 0.2$ (left) and $e_0 = 0.4$
    (right).}
  \end{center}
\end{figure}

Figure~\ref{sample_epoch} shows the phase space in mean semi-major axis and
longitude for $A/m~=~10~\mathrm{m^2/kg}$ and fixed values of the 
initial conditions, 
namely $e_0 = 0.1$, $i_0 = 0.004$~rad, $\Omega_0 = \omega_0 = 0$~rad. The
differences between the two graphs only depends on the initial {\it
time at epoch} parameter $t_0$. We could actually expect that
different initial times at epoch, namely, different initial
ecliptic longitudes of the Sun $\lambda_\odot(0)$, will reveal a quite
rich collection of behaviors, depending on the different states with
respect to the before-mentioned {\it eccentricity equilibrium}. 
Actually, assuming an initial time at epoch of 21 December 2001,
we see clearly that the phase space is filled by a large number of chaotic 
orbits (Figure~\ref{sample_epoch}, left). On the contrary, starting with an
initial time at epoch of 21 March 2000, that is adopting a Sun pointing
longitude of perigee 
($\lambda_\odot(0) = 0$~rad), 
%the values reached by the \texttt{MEGNO}
%tend to be smaller associated with significantly narrower chaotic region 
the \texttt{MEGNO} values tend to be smaller and associated with 
significantly narrower chaotic regions, 
always located close to the separatrices
(Figure~\ref{sample_epoch}, right). In the latter case, the eccentricity 
presents only small yearly variations due to the proximity of the
theoretical equilibrium.
%-------------------------------------------------------------------------------
\\
Therefore, these results seem to suggest that
high amplitude variations of the eccentricity increase considerably
the extension of chaotic regions close to the separatrices and, 
conversely, small eccentricity variations seem to minimize considerably 
the extent of chaotic regions. To justify this assumption, we
performed a dedicated numerical simulation with the same set of parameters used
in the one reported in Figure~\ref{sample_epoch}, but considering higher values
of the initial eccentricity. The results are reported in
Figure~\ref{higher_eccentricities}, the 
chosen time at epoch is $21$~December 2000 and the initial eccentricities
are, $e_0 = 0.2$ (left panel) and $e_0 = 0.4$ (right panel). In the latter
case,  
the huge variations of the perigee altitude, induced by the large
variations of the eccentricity as well as by the variations of the 
semi-major axis, leads to even more complicated dynamics. These results
thus confirm the importance of the initial eccentricity in the appearance of
chaos. 

%\newpage
%-----------------------------------------------------------------------------
\section{Secondary resonances}\label{secondary_resonances_section}
%---------------------------------------------------------------------------
It is worth noting that inspecting
Figures~\ref{sample_various_asm_transf}, \ref{sample_epoch} and
\ref{higher_eccentricities} we clearly note the presence of some
additional patterns located on both sides of the separatrices in the
phase space. These never seen before regions, hence unexplained so far, are
actually characterized by very low \texttt{MEGNO}
values. Indeed, this observation underlines the fact that the dynamics
of high area-to-mass ratio space debris is even more intricate than
expected. In the following two sections we will provide some
numerical results and an analytical theory, based on a simplified model,
to better understand such zones.

\subsection{Numerical investigations}\label{numericalinvest}
We followed a large set of near-geosynchronous space debris, related
to an extremely large set of initial conditions taken on both sides of
the pendulum-like pattern, and for each one of the $72\, 000$ orbits
we computed the related \texttt{MEGNO} indicator. The initial
conditions have been fixed by a mean longitude grid of 1$^\circ$,
spanning 360$^\circ$, and a semi-major axis grid of 1~km, spanning the
$42\,164\pm100$~km range, while the remaining orbit parameters and time 
at epoch are the same as in Figure~\ref{sample_various_asm}. 
Moreover, as in the previous extended analyses, the model of 
forces also includes the central body attraction, the 
second degree and order harmonics
$J_2,C_{22}$ and $S_{22}$ as well as the combined attractions of the
Sun and the Moon. The perturbing effects of direct solar radiation
pressure are also taken into account for a high area-to-mass ratio
fixed at $10~\mathrm{m^2/kg}$.\\

\begin{figure}[!htbp]
  \begin{center}
    \begin{tabular}{c}
      \includegraphics[width=.85\textwidth]{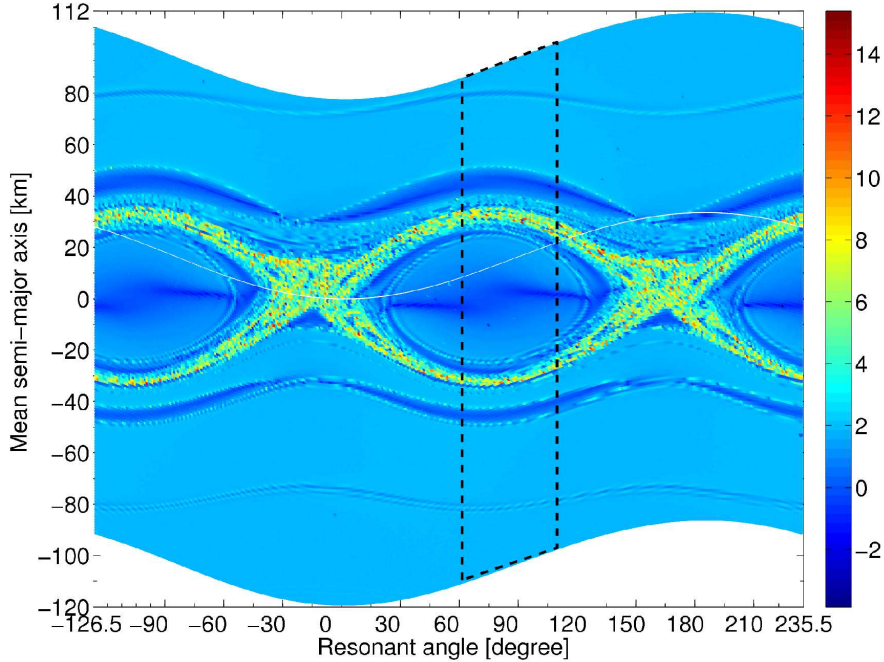}
      %    \\
      %    \includegraphics[width=.85\textwidth]{All_polar.eps}
    \end{tabular}
  \end{center} 
  \caption{\label{secondary_resonances} \texttt{MEGNO} computed as
    a function of initial mean longitude $\lambda_0$ and semi-major
    axis $a_0$. %[Top] Enlargement of the complete phase space
    %   phase. [Bottom] Enlargement of the complete phase space phase
    %   shown in polar coordinates. 
    The equations of motion include the
    central body attraction, the second degree and order harmonics
    $J_2,C_{22}$ and $S_{22}$ as well as the luni-solar
    perturbations. The mean longitude grid is $1^{\circ}$ and the
    semi-major axis grid is $1$~km, spanning the $42164 \pm 100$~km
    range. The initial conditions are $e_0 = 0.002$, $i_0 = 0.004$~rad 
    and $\Omega_0 = \omega_0 = 0$~rad. The area-to-mass ratio is 
    $10~\mathrm{m^2/kg}$. Time at epoch is 25~January~1991.}
\end{figure}

The results are reported in Figure~\ref{secondary_resonances}, which 
is nothing but an extensive enlargement of the phase space presented in
Figure~\ref{sample_various_asm}~(bottom, left). This phase space
widening clearly underlines the before-mentioned additional structures
located at $\pm$~40~km on each side of the resonant area. Furthermore,
besides these patterns, what is of special interest is that this
Figure also brings to the light supplementary structures located at
approximately 80~km on both sides of the main resonance, suggesting
that the phase space is actually foliated by a larger set of secondary
structures. Moreover, the width of these additional patterns and the 
numerical values of the \texttt{MEGNO} both seem to be directly 
related to the inverse of the distance with respect to the resonant area.
\\
%%%%%%%%%%%%%%%%%%%%%%%%%%%%%%%%%%%%%%%%%%%%%%%%%%%%%%%%%%%%%%%%%%%%%%%%%%% 

In addition, we also performed a set of similar numerical
investigations, in order to distinguish qualitatively the relative
relevance of some parameters such as the initial mean eccentricity, 
the value of the area-to-mass ratio, as well as the importance of the
1:1 resonance and of the third-body perturbations in the occurrence of
such secondary structures. Even though these results are not presented
here in detail, we can draw the following preliminary conclusions: the
second order harmonic $J_2$, as well as the third-body perturbations, do
not seem to be really relevant and crucial in the appearance of these
additional patterns. In other words, the unexpected patterns occur
only when taking into account the combined effects of both the second
order and degree harmonic and direct radiation pressure. As a
matter of fact, the extended numerical investigations performed in
Figure~\ref{sample_various_asm}~(top, left), or similarly those
shown in \citet{breiter2005}, also present these structures, even
though they are difficult to perceive. Actually, the extension and 
chaoticity indicator of the secondary patterns seem to be directly 
proportional to the area-to-mass ratio value or, equivalently 
directly proportional to the mean value of the eccentricity.
\begin{figure}[!t]
  \begin{center}
    \includegraphics[width=\textwidth]{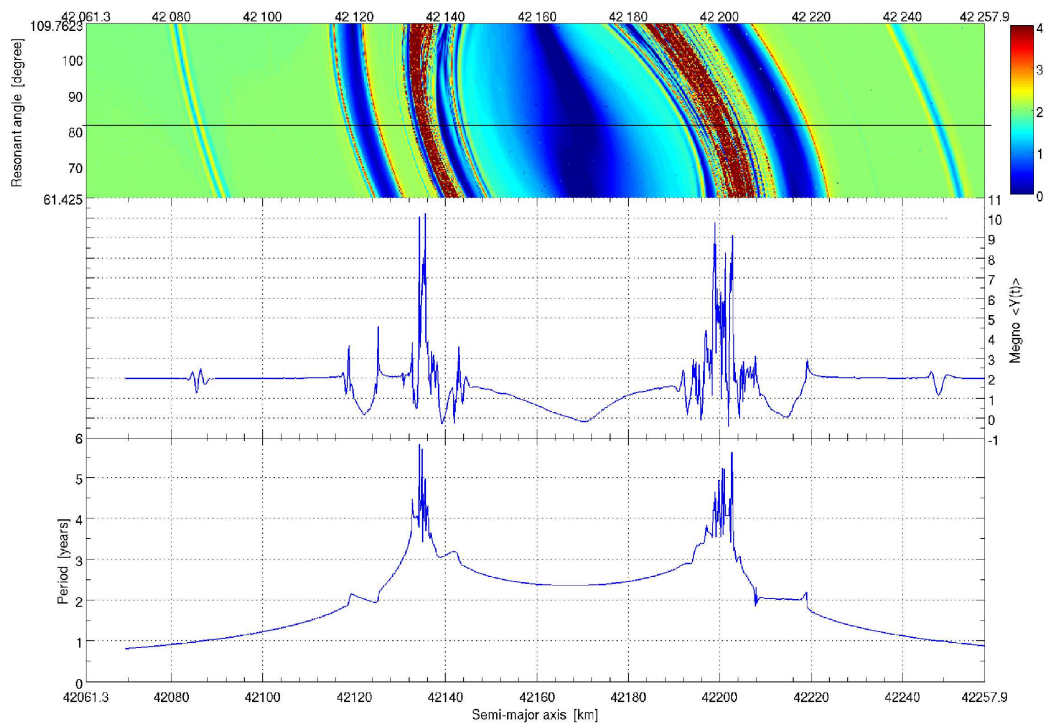} 
  \end{center}
  \caption{\label{frequency_analysis}Blow-up of the phase space with
    the specification of a {\it resonant angle section} (horizontal
    black solid line), that is the set of orbits having the same
    (osculating) initial resonant angle value, near the first stable
    equilibrium point, namely $\sigma_0^{section} = 81.67^{\circ}$ (top
    panel). Evolution of \texttt{MEGNO} with respect to the
    initial semi-major axis $a_0$ for the specified section (middle
    panel). The fundamental period of $\sigma$ with respect to the
    initial semi-major axis $a_0$, computed by means of frequency
    analysis for the specified section (bottom panel). The estimation
    of the periods are made over a 20~years period of time.}
\end{figure}\\

To get even more concluding results, we considered a blow-up of the
phase space (dashed line rectangle in
Figure~\ref{secondary_resonances}) with really high resolution 
sampling (150~meters in the semi-major axis~$a$ and
$0.3^\circ$ in the resonant angle~$\sigma$). 
Figure~\ref{frequency_analysis}~(top) shows this phase
space widening wherein we defined a so-called {\it resonant angle
section} (horizontal black solid line), that is the subset of orbits
having the same initial resonant angle value. This resonant angle
section spans the complete range in semi-major axis and passes close to
the stable equilibrium point. For each orbit defined on this section,
we computed the \texttt{MEGNO} indicator and in
Figure~\ref{frequency_analysis}~(middle) we report this value at the
end of the simulation as a function of the semi-major axis.

To double check our results, we performed a frequency analysis
%investigation (see \citealt{Laskar2,Laskar}, and \citealt{noyelles}) aimed to
investigation (see Laskar, 1990 and 1995, and Noyelles et al., 2008) aimed to
study the behavior of the proper frequency of the resonant angle
$\sigma$, whose results are reported in
Figure~\ref{frequency_analysis}~(bottom). Here one can clearly notice 
the distinctive characteristics regarding the well-known 1:1 resonance
between the mean longitude and the sidereal time. Indeed, both 
\texttt{MEGNO} and the fundamental period show distinctively a minimum
close to the stable equilibrium point. In this case, as previously
mentioned in Section~\ref{validation_section}, \texttt{MEGNO}
should slowly converge to $\overline{Y}(t) = 2$ everywhere, except at
the equilibrium point where the limit value is $\overline{Y}(t) = 0$,
that's why, using a finite integration time, we obtain such V-shaped
curve, close to 0 in the center of the resonance and to 2 on the
borders. It is also worth noting that the fundamental period of
$\sigma$ is reported to be close to 2.25~years, which is in good
agreement with the well-known 818~days libration period of a typical
uncontrolled near-geosynchronous object. Near the separatrices, 
\texttt{MEGNO} clearly presents some obvious high values which
confirms the presence of chaotic orbits. Here, the fundamental period
reaches significant values and, as a matter of fact, is not well
determined, once again supporting the result of the existence of a
chaotic zone.
\begin{figure}[!t]
  \begin{center}
    \includegraphics[width=\textwidth]{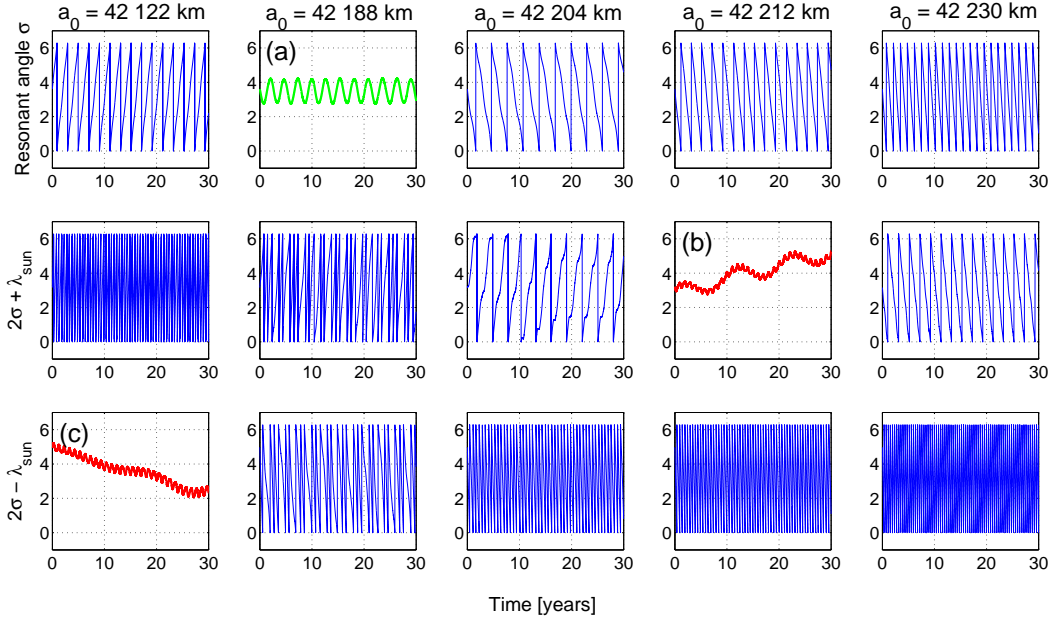}
  \end{center}
  \caption{\label{secresosig}Time evolution of the angles $\sigma$,
    $2\sigma+\lambda_{\odot}$ and $2\sigma-\lambda_{\odot}$ (in radians) for
    several semi-major axes. In the lower major secondary resonance,
    $a_0=42\,122$~km. In the eye of the principal resonance,
    $a_0=42\,188$~km. Between the primary resonance and the upper
    secondary resonance, $a_0=42\,204$~km. Inside the upper major
    secondary resonance, $a_0=42\,212$~km. Outside the upper major
    secondary resonance, $a_0=42\,230$~km.}
\end{figure}\\
%% Differences between mean and osculating
%% semi-major axis for three horizontal sampling (fixed osculating
%% semi-major axes). The first difference is related to a semi-major axis
%% sampling taken above the libration region, the second is related to a
%% semi-major axis sampling which crosses the libration region and
%% finally, the third sampling is taken below this region.

Moreover, the use of frequency analysis allows us to support strongly 
the hypothesis that the additional patterns
are actually related to \textbf{\emph{secondary resonances}}. Indeed, 
if we look at the evolution of the fundamental period with respect to
the semi-major axis, it is clear that the so-called secondary
resonances are associated, regarding the angle $\sigma$, with periods
which are commensurate with 1~year. More precisely, the major
secondary resonances, located at approximately 40~km on both sides of
the pendulum-like pattern, are related to a 2~years fundamental period
of $\sigma$. Concerning the farther patterns located at $\pm$80~km,
the fundamental period of $\sigma$ turns out to be very close to
1~year. As a consequence, we can presumably assume that these
secondary resonances are actually related to a
commensurability between $\sigma$ and the 1~year period angle
$\lambda_\odot$, that is the ecliptic longitude of the Sun.

To justify this assumption, we focused our attention to the
major secondary resonances located at $\pm$40~km on both sides of the
pendulum-like pattern, considering the time evolution of various
linear combinations of $\sigma$ and $\lambda_\odot$. For this
purpose, we considered various initial semi-major axes in the phase
space. The results are shown in Figure~\ref{secresosig}. At first
glance, it is apparent that three propagations stand apart from the 
others. In the first row of Figure~\ref{secresosig}, that is regarding
the evolution of the resonant angle $\sigma$, we clearly identify the
well-known characteristics related to the primary resonance. In
particular, in Figure~\ref{secresosig}a, that is when considering an
initial semi-major axis inside the primary resonant
($a_0=$42\,188~km), $\sigma$ shows the well-known long-period 
libration (2.25~years), whereas $\sigma$ circulates outside this
region. Furthermore, what is of special interest is the time evolution
of both $2\sigma + \lambda_\odot$ and $2\sigma - \lambda_\odot$, shown
in the second and third row, respectively. It is clear that most of
the time these angles show a circulation regime. However,
when considering an initial semi-major axis inside the major lower
secondary resonance for $2\sigma - \lambda_\odot$ or, similarly inside
the major upper secondary resonance for $2\sigma + \lambda_\odot$,
both these angles show a significant long-term evolution
(Figure~\ref{secresosig}b and \ref{secresosig}c).
%%   Indeed the tableband appear
%% when the value of resonant angle is equal to $2$ and $1$ (visible when
%% you zoom). We deal with secondary resonance $2\sigma:1\lambda$ and
%% $1\sigma:1\lambda$ (lower eye) and $-2\sigma:1\lambda$ and
%% $-1\sigma:1\lambda$ (upper eye).\\
%% Making sure we draw in the Figure 15 most plot with the same mean anomaly
%% ($3.60301$ rad) but with a different semi-major axis : one in the lower
%% secondary resonance, one in the eye, one between the eye and the first upper
%% secondary resonance, one in this resonance and one upper this. For each
%% semi-major axis we plot the resonant angle $\sigma$, the NEW-resonant-angle
%% $2\sigma+\lambda_{\odot}$ and $2\sigma-\lambda_{\odot}$. We notice that this
%% angle is in fast circulation all over exept in the secondary resonance that we
%% have select that confirm our intuition.
%% In the next part, we present/introduce an analytical explication of
%% this secondary resonance and we confirm/corroborate our investigation by a
%% numerical test in the last part.
%%%%%%%%%%%%%%%%%%%%%%%%%%%%%%%%%%%%%%%%%%%%%%%%%%%%%%%%%%%%%%%%%%%%%%%%%%%%%%%%%%%%

\subsection{Analytical investigation -- simplified model}
The presence and the location of these secondary
resonances can be studied using an appropriate simplified model. Hence we model
the averaged geostationary motion by a pendulum-like system, given by its
Hamiltonian formulation~\citep{valk07a} up to order $e^2$ in 
the series expansion
\begin{equation*} 
\mathcal{H} = - \frac{\mu^2}{2 L^2} - \dot \theta L + 3\,\frac{\mu^4}{L^6} \;  R_{e}^2  \;  \left(1 - \frac{5}{2} \, e^2\right) \; S_{2200}(\Omega,\omega,M,\theta)\,, 
\end{equation*}
where 
$$
\displaystyle L = \sqrt{\mu a }\, \qquad \text{and} \quad
\displaystyle S_{2200}(\Omega,\omega,M,\theta) =C_{22} \cos\,2\sigma +
S_{22} \sin\,2\sigma\,.
$$ 
In the context of direct solar radiation
pressure, we can introduce the factor $\mathcal{Z}$ proportional to
$A/m$ through the eccentricity $e$ (for further details, we refer to
the averaged simplified analytical model developed in 
%\citealt{valk07b}). As a first approximation, the time evolution of both
Valk et al., 2008). As a first approximation, the time evolution of both 
the eccentricity $e$ and the longitude of perigee $\varpi$ were found
to be (neglecting the obliquity of the Earth with respect to the
ecliptic)
\begin{equation*}
%\baselinestretch{1.4}
\setstretch{1.4}
\begin{array}{lcl} 
\displaystyle e \cos \varpi &=& \displaystyle \frac{\mathcal{Z}}{L\,n_\odot} \cos \lambda_\odot + \alpha_0\,, \\ 
\displaystyle e \sin \varpi &=& \displaystyle \frac{\mathcal{Z}}{L\,n_\odot} \sin \lambda_\odot - \beta_0\,,
\end{array}
\end{equation*}
which introduces $\lambda_\odot$ in the Hamiltonian. The quantity
$n_\odot$ is the mean motion of the Sun and both $\alpha_0$ and 
$\beta_0$ are related to the initial conditions with respect to the
eccentricity and the longitude of perigee). The resulting Hamiltonian 
takes the generic form
\begin{equation*}
\mathcal{H} =  - \frac{\mu^2}{2 L^2} - \dot \theta L + \frac{F}{L^6} \;  \cos \, (2 \sigma- 2 \sigma_0) 
- \frac{G}{L^6} \; 2 \cos \, (2 \sigma- 2 \sigma_0) \,
\cos\, (\lambda_\odot + \delta)\,,
\end{equation*}
 where $\delta$, $F$, $G$, $\sigma_0$ are constants.  A suitable
 transformation is then necessary to introduce action-angle variables
 $(\psi, J)$ in the libration and in the circulation region of the
 double pendulum, in such a way any trajectory of the double pendulum
 is characterized by a constant action $J$ and a corresponding
 constant frequency $\dot \psi$. Rewriting the perturbed system
 (because of $\lambda_\odot$ terms) by means of these new variables
 and then using the expansions in Bessel functions, we could isolate
 any resonance of the type $k \psi \pm \lambda_\odot$ in the
 circulation region, for any $|k|$, and in the libration region, for
 $|k| \geq 3$, which corresponds to our frequency analysis. This
 analysis is surely promising, but it is outside the goals of this
 paper. Further investigations will be detailed in a
 forthcoming publication \citep{Lemaitre2009}.

\section{Conclusions}
The predictability of the trajectory high area-to-mass ratio space 
debris located near the geosynchronous region was investigated 
by means of a recent variational chaos indicator called \texttt{MEGNO}. 
Thanks to this technique, we clearly identified the regular 
(stable) and irregular (chaotic) orbits. This efficient method allowed 
us to obtain a clear picture of the phase space, hence showing that chaotic
regions can be particularly relevant, especially for very high area-to-mass
ratio objects. Moreover, we discussed the importance of both the initial
eccentricity and time at epoch in the appearance of chaos.

Finally, we brought to the fore a relevant class of
additional unexpected patterns which were identified as secondary
resonances, that were numerically studied by
means of both the \texttt{MEGNO} criterion and frequency map analysis, to
eventually conclude that they involve
commensurabilities between the primary resonant angle and
the ecliptic longitude of the Sun. We also presented an analytical scheme that
could explain their existence. It will be the subject of further work.

%++++++++++++++++++++++++++++++++++++++++++++++++++++++++++++++++++++++++
\section*{Acknowledgement}
The authors thank S. Breiter for helpful discussions about both the
\texttt{MEGNO} criterion as well as numerical issues which led to
substantial improvement of the present paper, also providing some
useful references. Finally, we are grateful for the opportunity to use
the frequency analysis tools developed by B.~Noyelles and A.~Vienne. 
The authors also warmly thank the two referees for their suggestion that 
allowed us to strongly improved the paper.

%\bibliographystyle{elsart-harv}
%\bibliographystyle{elsart-num}
%\bibliographystyle{elsart-num-names}
%\bibliographystyle{elsart-num-sort}
%\bibliography{biblioPapier}

\begin{thebibliography}{28}
\expandafter\ifx\csname natexlab\endcsname\relax\def\natexlab#1{#1}\fi
\expandafter\ifx\csname url\endcsname\relax
  \def\url#1{\texttt{#1}}\fi
\expandafter\ifx\csname urlprefix\endcsname\relax\def\urlprefix{URL }\fi

\bibitem[{{Anselmo} and {Pardini}(2005)}]{anselmo05}
{Anselmo}, L., {Pardini}, C., Orbital evolution of geosynchronous objects
  with high area-to-mass ratios. In: Danesy, D. (Ed.), Proceedings of the
  Fourth European Conference on Space Debris, (ESA SP-587). ESA Publications
  Division, Noordwijk, The Netherlands, pp. 279--284, 2005.

\bibitem[{{Barrio} et~al.(2007){Barrio}, {Borczyk}, and {Breiter}}]{barrio2007}
{Barrio}, R., {Borczyk}, W., {Breiter}, S., Spurious structures in chaos
  indicators maps. In press, Chaos, Solitons \&
  Fractals, doi:10.1016/j.chaos.2007.09.084, 2007.

\bibitem[{{Benettin} et~al.(1980){Benettin}, {Galgani}, {Giorgilli}, and
  {Strelcyn}}]{benettin80a}
  {Benettin}, G., {Galgani}, L., {Giorgilli}, A., et al.,  %{Strelcyn}, J.-M.,
  Lyapunov characteristic exponents for smooth dynamical systems and for
  hamiltonian systems; a method for computing all of them. Part 1: Theory.
  Meccanica 15, 9--20, 1980

\bibitem[{{Breiter} et~al.(2005){Breiter}, {Wytrzyszczak}, and
  {Melendo}}]{breiter2005}
{Breiter}, S., {Wytrzyszczak}, I., {Melendo}, B., Long-term
  predictability of orbits around the geosynchronous altitude. Advances in
  Space Research 35, 1313--1317, 2005.

\bibitem[{{Bulirsh} and {Stoer}(1966)}]{bulirsh-stoera}
{Bulirsh}, R., {Stoer}, J., Numerical treatment of ordinary
  differential equations by extrapolation methods. Numerische Mathematik 8,
  1--13, , March 1966.

\bibitem[{{Chao}(2006)}]{chao06}
{Chao}, C.~C., Analytical investigation of {GEO} debris with high
  area-to-mass ratio, AIAA Paper No. AIAA-2006-6514, Presented at the 2006
  AIAA/AAS Astrodynamics Specialist Conference, Keystone, Colorado,
  August 2006.

\bibitem[{{Cincotta} et~al.(2003){Cincotta}, {Giordano}, and
  {Sim\'{o}}}]{cincotta2003}
{Cincotta}, P.~M., {Giordano}, C.~M., {Sim\'{o}}, C., Phase space
  structure of multi-dimensional systems by means of the mean exponential
  growth factor of nearby orbits. Physica D 182, 151--178, 2003.

\bibitem[{{Cincotta} and {Sim\'{o}}(2000)}]{cincotta2000}
{Cincotta}, P.~M., {Sim\'{o}}, C., Simple tools to study global dynamics
  in non-axisymmetric galactic potentials - {I}. Astronomy and Astrophysics,
  Supplement 147, 205--228, 2000.

\bibitem[{{Cunningham}(1970)}]{cunningham70}
{Cunningham}, L.~E., On the computation of the spherical harmonics terms
  needed during the numerical integration of the orbital motion of an
  artificial satellite. Celestial Mechanics 2, 207--216, 1970.

\bibitem[{{Deprit}(1969)}]{deprit69}
{Deprit}, A., Canonical transformations depending on a small parameter.
  Celestial Mechanics 1, 12--30, 1969.

\bibitem[{{Go\'{z}dziewski} et~al.(2001){Go\'{z}dziewski}, {Bois},
  {Maciejewski}, and {Kiseleva-Eggleton}}]{gozdziewski2001}
  {Go\'{z}dziewski}, K., {Bois}, E., {Maciejewski}, A.~J., et al., %{Kiseleva-Eggleton},  L., 
  Global dynamics of planetary systems with the \texttt{MEGNO}
  criterion. Astronomy and Astrophysics 378, 569--586, 2001.

\bibitem[{{Go\'{z}dziewski} et~al.(2008){Go\'{z}dziewski}, {Breiter}, and
  {Borczyk}}]{gozdziewski2007}
{Go\'{z}dziewski}, K., {Breiter}, S., {Borczyk}, W., The long-term
  stability of extrasolar system {HD}37154. numerical study of resonance 
  effects. Monthly Notices of the RAS 383, 989--999, 2008.

\bibitem[{{Henrard}(1970)}]{henrard70}
{Henrard}, J., On a perturbation theory using lie transforms. Celestial
  Mechanics 3, 107--120, 1970.

\bibitem[{{Laskar}(1990)}]{Laskar2}
{Laskar}, J., The chaotic motion of the solar system: a numerical
  estimate of the size of the chaotic zones. Icarus 88, 266--291, 1990.

\bibitem[{{Laskar}(1995)}]{Laskar}
{Laskar}, J., Introduction to frequency map analysis. In:
  Proceedings of 3DHAM95 NATO Advanced Institute. Vol. 533. S'Agaro, 
  134--150, June 1995.

\bibitem[{{Lema\^{i}tre} et~al.(submitted for publication){Lema\^{i}tre}, {Delsate}, and  {Valk}}]{Lemaitre2009}
  {Lema\^{i}tre}, A., {Delsate}, N., {Valk}, S., 
  A web of secondary resonances for large $A/m$ geostationary debris.
  Submitted to Celestial Mechanics and Dynamical Astronomy, 2009.

\bibitem[{{Lemoine} et~al.(1987){Lemoine}, {Kenyon}, {Factor}, {Trimmer},
  {Pavlis}, {Chinn}, {Cox}, {Klosko}, {Luthcke}, {Torrence}, {Wang},
  {Williamson}, {Pavlis}, {Rapp}, and {Olson}}]{lemoine87}
{Lemoine}, F.~G., {Kenyon}, S.~C., {Factor}, J.~K., et al., 
%{Trimmer}, R., {Pavlis},
%  N.~K., {Chinn}, D.~S., {Cox}, C.~M., {Klosko}, S.~M., {Luthcke}, S.~B.,
%  {Torrence}, M.~H., {Wang}, Y.~M., {Williamson}, R.~G., {Pavlis}, E.~C.,
%  {Rapp}, R.~H., {Olson}, T.~R., 
  The development of the joint nasa gsfc
  and nima geopotential model {EGM}96. Tech. rep., NASA, {TP}-1998-206861, 1987.

\bibitem[{{Liou} and {Weaver}(2004)}]{liouweaver04}
{Liou}, J.-C., {Weaver}, J.~K., Orbital evolution of {GEO} debris with very
  high area-to-mass ratios. The Orbital Quarterly News 8 issue 3, {T}he NASA
  Orbital Debris Program Office, 2004.

\bibitem[{{Liou} and {Weaver}(2005)}]{liouweaver05}
{Liou}, J.-C., {Weaver}, J.~K., Orbital dynamics of high area-to-mass
  ratio debris and their distribution in the geosynchronous region. In: Danesy,
  D. (Ed.), Proceedings of the Fourth European Conference on Space Debris (ESA
  SP-589). ESA Publications Division, Noordwijk, The Netherlands,
  pp. 285--290, 2005.

\bibitem[{{Noyelles} et~al.(2008){Noyelles}, {Lema\^{i}tre}, and
  {Vienne}}]{noyelles}
{Noyelles}, B., {Lema\^{i}tre}, A., {Vienne}, A., Titan's rotation. A
  $3$-dimensional theory. Astronomy and Astrophysics 475, 959--970, 2008.

\bibitem[{{Schildknecht} et~al.(2005){Schildknecht}, {Musci}, {Flury},
  {Kuusela}, {de Leon}, and {de Fatima Dominguez Palmero}}]{schildknecht05}
  {Schildknecht}, T., {Musci}, R., {Flury}, W., et al., 
%{Kuusela}, J., {de Leon}, J., {de Fatima Dominguez Palmero}, L., 
  Optical observations of space debris in
  high-altitude orbits. In: Danesy, D. (Ed.), Proceedings of the Fourth
  European Conference on Space Debris, ESA SP-587. ESA Publications Division,
  Noordwijk, The Netherlands, pp. 113--118, 2005.

\bibitem[{{Schildknecht} et~al.(2004){Schildknecht}, {Musci}, {Ploner},
  {Beutler}, {Flury}, {Kuusela}, {Leon Cruz}, and {de Fatima Dominguez
  Palmero}}]{schildknecht04}
  {Schildknecht}, T., {Musci}, R., {Ploner}, M., et al., 
  %{Beutler}, G., {Flury}, W., {Kuusela}, J., {Leon Cruz}, J., {de Fatima Dominguez Palmero}, L.,
  Optical observations of space debris in {GEO} and in highly-eccentric orbits.
  Advances in Space Research 34, 901--911, 2004.

\bibitem[{{Standish}(1998)}]{standish98}
{Standish}, E.~M., {JPL} planetary and lunar ephemeris,
  {DE}405{/}{LE}405. {JPL} Interoffice Memorandum IOM 312.D-98-048, August 1998.

\bibitem[{{Stoer} and {Bulirsch}(1980)}]{bulirsh-stoerb}
{Stoer}, J., {Bulirsch}, R., Introduction to numerical analysis.
  Springer-Verlag, New York, 1980.

\bibitem[{{Valk et~al.}(2008a)}]{valk07b}
{Valk}, S., {Lema\^{i}tre}, A., {Anselmo}, L., Analytical and
  semi-analytical investigations of geosynchronous space debris with high
  area-to-mass ratios influenced by solar radiation pressure. Advances in Space
  Research 41, 1077--1090, 2008a.

\bibitem[{{Valk} and {Lema\^{i}tre}(2008b)}]{valk08}
{Valk}, S., {Lema\^{i}tre}, A., Semi-analytical investigations of high
  area-to-mass ratio geosynchronous space debris including earth's shadowing
  effects. Advances in Space Research 42, Issue 8, 1429--1443, 2008b.

\bibitem[{{Valk et~al.}(2008c)}]{valk07a}
{Valk}, S., {Lema\^{i}tre}, A., {Deleflie}, F., 
  Semi-analytical theory of mean orbital motion for geosynchronous space debris
  under gravitational influence. In press, Advances in Space Research, 
  doi: 10.1016/j.asr.2008.12.015, 2008c.

\bibitem[{{Wisdom}(1983)}]{wisdom83}
{Wisdom}, J., Chaotic behavior and the origin of the 3/1 {K}irkwood gap.
  Icarus 56, 51--74, 1983.

\bibitem[{{Wytrzyszczak} et~al.(2007){Wytrzyszczak}, {Breiter}, and
  {Borczyk}}]{wytrzyszczak2007}
{Wytrzyszczak}, I., {Breiter}, S., {Borczyk}, W., Regular and chaotic
  motion of high altitude satellites. Advances in Space Research 40, 134--142,
  2007.

\end{thebibliography}

\end{document}